\documentclass[aps,prd,twocolumn,showpacs,floats,floatfix,letterpaper,nofootinbib,superscriptaddress,]{revtex4-1}

\usepackage{amssymb,amsmath,latexsym,mathrsfs}
\usepackage{graphicx,subfigure}
\usepackage{epsfig}
\usepackage{varioref,xr-hyper}
\usepackage{color}
\usepackage{multirow}
\usepackage{array}
\begin{document}
	
\title{Understanding the Dyer-Roeder approximation as a consequence of local cancellations of projected shear and expansion rate fluctuations}
\author{S. M. Koksbang}
\email{koksbang@cp3.sdu.dk}
\affiliation{$CP^3$-Origins, University of Southern Denmark, Campusvej 55, DK-5230 Odense M, Denmark}
	
\begin{abstract}
It is shown with several concrete examples that the Dyer-Roeder approximation is valid in spacetimes which fulfill the condition that fluctuations in the expansion rate along a light ray locally cancels with the shear contribution to the redshift. This is the case for standard cosmological scenarios including perturbed FLRW spacetimes, N-body simulations and Swiss-cheese models. With another concrete example it is then illustrated that it is possible to construct statistically homogeneous and effectively statistically isotropic cosmological models which do not fulfill the condition. In this case, the Dyer-Roeder approximation is invalid. Instead, the mean redshift-distance relation can be described using a relation based on the spatial averages of the transparent part of the spacetime.
\end{abstract}

\maketitle
	
\section{Introduction}
By studying light propagation in e.g. N-body simulations it has been found that narrow beams from point sources such as supernovae typically travel through a mean mass density significantly lower than the cosmic mean - at least at low redshifts \cite{misinterp,cc2,Bolejko_RicciShear,HolzWald}. The most well-known method for taking this into account when describing the redshift-distance relation is the Dyer-Roeder approximation \cite{DR1,DR2,Zeldovich}. This approximation is based on the idea that a fraction of the matter in the Universe is localized in clumps that (narrow) light beams rarely meet such that the mean density sampled by these light beams is less than the spatial mean. More specifically, three assumptions are made in the Dyer-Roeder approximation namely that 1) the redshift along narrow beams is the same as that of the background FLRW model, 2) Weyl lensing can be neglected, and 3) the Ricci lensing can be modified simply by multiplying a constant $\alpha$ onto the density in order to take the mean underdensity along the narrow beams into account. This results in a redshift-distance relation which can be written as ($c = 1$ throughout)
\begin{align}\label{eq:DR}
	\frac{d^2D_A}{dz^2} + \left(\frac{2}{1+z}+\frac{d\ln H}{dz} \right) \frac{dD_A}{dz} = -\frac{4\pi G\rho}{H^2\left( 1+z\right)^2} \alpha D_A.
\end{align}
By analyzing supernova data with this redshift-distance relation, limits obtained for $\alpha$ so far indicate $0.4\lesssim \alpha$ \cite{alphavalue1,alphavalue2,alphavalue3,alphavalue4,alphavalue5,alphavalue6,alphavalue7,alphavalue8}, although the exact constraints naturally depend much on what other restrictions are introduced including e.g. on the dark energy equation-of-state. If a flat $\Lambda$CDM model is assumed, the value of $\alpha\approx 1$ is obtained \cite{alphavalue5}. A value of $\alpha \approx 1$ is also indicated by the fact that results based on standard $\Lambda$CDM interpretations of supernova data is overall consistent with observations based on much wider beams (e.g. BAO and CMB data) which are expected to sample a mass density corresponding to the cosmic mean.
\newline\indent
From a theoretical point of view, an observational value of $\alpha\approx 1$ would at this point be somewhat surprising since theoretical investigations of e.g. N-body simulations indicate that narrow beams should typically have $\alpha$ significantly smaller than 1 (but see e.g. \cite{flux1,flux2,flux3,flux4} for a possible, at least partial, explanation based on flux-averaging). One may note though, that while typical narrow light beams are theoretically expected to experience an underdense universe, a small number of special rays will experience a compensating significant lensing when passing close by structures of high density. The resulting mean redshift-distance relation when averaging over the entire sky is therefore expected to be well described by that of the FLRW background even when considering narrow beams (see e.g. \cite{Weinberg} as well as the more recent, detailed considerations of \cite{dowecare,Kaiser,KibbeLieu}). However, our supernovae catalogs do not cover the entire sky and additionally one may worry that the special rays, if even observed, would be rejected as outliers (see e.g. \cite{teppo} for a discussion). It is therefore far from clear how $\alpha\approx 1$ should come about. An obvious related question is whether real supernova light beams actually do typically experience an underdense spacetime but that this may go unnoticed when interpreting data with the Dyer-Roeder approximation because this approximation is insufficient for describing the effect. Indeed, the Dyer-Roeder approximation is a conjecture with underlying assumptions which have not all been justified theoretically. Specifically, while the assumptions of neglecting Weyl lensing and reducing the Ricci lensing by a factor $\alpha$ is reasonable for light rays traveling mostly in underdense regions far from high-density structures, the assumption that the redshift should be given by the background expression does not seem to have a clear theoretical justification.
\newline\newline
Since the Dyer-Roeder approximation is largely a conjecture, the literature naturally contains studies into possible modifications of the approximation. Such introductions e.g. include introducing a redshift dependence of $\alpha$ which certainly has some theoretical justification \cite{Bolejko_letter, Santos_DR, Linder_DR, review, teppo, Mortsell}. At a more fundamental level, it has also been questioned whether it is valid not to take into account any effect it may have on the mean expansion rate and on the redshift along the light beams that these sample an underdense universe \cite{misinterp, cc2,EhlersSchneider_DRcritique, teppo}. In \cite{DR_SMK} it was even argued that it {\em a priori} seems more reasonable to expect that the mean redshift-distance relation follows the prescription given in \cite{syksy1,syksy2} where the redshift-distance relation is based on spatially averaged quantities, yielding
\begin{align}\label{eq:av}
\begin{split}
\frac{d^2D_A}{dz_{\rm av}^2} + \left(\frac{2}{1+z_{\rm av}}+\frac{d\ln  H_{\rm av} }{dz_{\rm av}} \right) \frac{dD_A}{dz_{\rm av}} = -\frac{4\pi G\rho_{\rm av} }{H_{\rm av} ^2\left( 1+z_{\rm av}\right)^2} D_A,
\end{split}
\end{align}
where $\rho_{\rm av}$ is the spatially averaged density field, and $H_{\rm av}$ a third of the averaged local expansion rate $\theta$ (with averages given by $X_{\rm av}:=\frac{\int_D XdV}{\int_D dV}$, where $dV$ is the proper infinitesimal volume element of the spatial hypersurfaces, and $D$ the spatial averaging domain). The mean redshift is in \cite{syksy1,syksy2} argued to be well approximated by
\begin{equation}\label{eq:syksy_z}
1+z_{\rm av} = \exp\left( \int_{t_e}^{t_o}dtH_{\rm av}\right) ,
\end{equation}
where $t_o$ is the time of observation and $t_e$ that of emission (the former usually taken to be present time, $t_0$).
\newline\indent
The redshift-distance relation in Eq. \ref{eq:av}, \ref{eq:syksy_z} was in \cite{syksy1,syksy2} derived for statistically spatially homogeneous and isotropic spacetimes with slowly evolving structures and without opaque regions. For narrow beams, the naive expectation would be that the spatial averaging domains used for computing the spatially averaged quantities in the above equations should simply exclude regions not probed by the light rays (such regions could be considered effectively opaque). The resulting relation differs from the Dyer-Roeder expression since the redshift is modified if there are (effective or actual) opaque regions.
\newline\indent
The prescription of Eq. \ref{eq:av}, \ref{eq:syksy_z} for computing a redshift-distance relation based on spatial averages has been found to yield an accurate approximation of the mean redshift-distance relation even in models that do not have explicit FLRW backgrounds \cite{tworegion_light,Hellaby_light}, but the models studied had no opaque regions. It was then shown in \cite{DR_SMK} that the redshift-distance relation based on spatial averages does {\em not} in general give the correct mean redshift-distance relation along light rays in Einstein-Straus models \cite{EinsteinStraus} which are Swiss-cheese models constructed by gluing Schwarzschild regions (with opaque centers) into an FLRW background. The Dyer-Roeder approximation was, on the other hand, found to give an excellent description of the mean redshift-distance relation along the light rays (in agreement with the results of \cite{Fleury}). It was found that the reason for the success of the Dyer-Roeder approximation is a delta-function contribution to the redshift which ensures that the redshift in these models will always be (nearly) identical to the background redshift after a light ray has traversed one or more entire structures.
\newline\indent
While there are indications that the same result could be obtained without introducing delta-functions by choosing another spacetime foliation (see e.g. \cite{Fleury}), it is interesting to understand exactly under what conditions one can expect the redshift to behave in such a way that it is always close to the background redshift, and whether such a  condition is necessary for the Dyer-Roeder approximation to be correct. Therefore, the work in \cite{DR_SMK} is here elaborated by considering the validity of the Dyer-Roeder approximation in three different types of inhomogeneous cosmological models and specifically studying under what conditions the redshift will be well approximated by the redshift of the background. In section \ref{sec:swisscheese}, Swiss-cheese models will be considered. Then, in section \ref{sec:nbody}, N-body simulations and perturbation theory is studied. Finally, in section \ref{sec:hellaby}, a more exotic model type is considered before concluding in section \ref{sec:conclusion}.

\section{Swiss-cheese models}\label{sec:swisscheese}
This section discusses the mean redshift-distance relation and specifically the mean redshift along light rays in Swiss cheese models.
\newline\newline
Swiss cheese models are based on FLRW backgrounds but have been made inhomogeneous by removing spherically symmetric patches of the FLRW background and substituting them by inhomogeneous models. If this substitution is done such that the Darmois junction conditions \cite{Darmois} are fulfilled, the resulting inhomogeneous model is an exact solution to the Einstein equations. The original version of the Swiss-cheese model, the Einstein-Straus model \cite{EinsteinStraus}, introduced Schwarzschild regions into an FLRW background. The light propagation qualities of the Einstein-Straus model has been studied in e.g. \cite{DR_SMK,Fleury,light_ES0,light_ES1,light_ES2,light_ES3,light_ES5}. More recently, Swiss-cheese models with Lemaitre-Tolman-Bondi (LTB) \cite{LTB1,LTB2,LTB3} and Szekeres \cite{Szekeres,Szekeres2} inhomogeneities have been extensively studied. It has become well-known that the mean redshift-distance relation in this type of Swiss-cheese models (without opaque regions) is well described by that of the background (see e.g. \cite{ltb_light_early} for one of the earlier papers on this topic). Swiss-cheese models with opaque regions have only been studied in terms of the Einstein-Straus model where the Dyer-Roeder approximation describes the mean redshift-distance relation well if the density used in the Ricci lensing term is given by the spatially averaged density of the transparent regions \cite{DR_SMK} or, equivalently, by the background density field multiplied by the volume fraction of transparent regions \cite{Fleury,light_ES5}.
\newline\newline
While it was in \cite{DR_SMK} found that the redshift along light rays in Einstein-Straus models is (nearly) equal to the background redshift only because of a delta-function contribution to the redshift, it was noted that a similar result would presumably be true for Swiss-cheese models based on LTB and Szekeres structures. That this is indeed correct is illustrated here by considering two concrete examples.
\newline\newline
In the subsections below, the LTB and Szekeres models are introduced, an appropriate light propagation formalism in the corresponding Swiss-cheese models is discussed and results from studying light propagation in Swiss-cheese models with LTB/Szekeres structures with opaque regions are presented.

\subsection{Model construction}
The Szekeres model is a family of exact solutions to the Einstein equations containing comoving dust and a cosmological constant. Its line element can be written as (subscripted commas indicate partial derivatives)
\begin{align}
\begin{split}
ds^{2} = -dt^{2} +\frac{\left(A_{,r}-A\frac{E_{,r}}{E}\right)^2}{\epsilon-k}dr^2 +\frac{A^2}{E^2}(dp^2+dq^2),
\end{split}
\end{align}
where $A = A(t,r)$, $E = E(r,p,q)$ and $k = k(r)$. $\epsilon$ is a constant that may take the values $\pm1, 0$. When $\epsilon = 1$, the model is known as the quasi-spherical Szekeres model. Only the quasi-spherical Szekeres model will be considered here. For a quasi-spherical Szekeres model $E$ can be written as
\begin{align}
	E = \frac{1}{2S}(p^2+q^2)-\frac{p P}{S} - \frac{qQ}{S}+\frac{P^2+Q^2+S^2}{2S},
\end{align}
where $P, Q$ and $S$ are arbitrary (though continuous) functions of $r$ with $S(r)\neq 0$. When $E_{,r} = 0$, the quasi-spherical Szekeres model reduces to the spherically symmetric LTB model. See e.g. \cite{wormhole} for more information on the Szekeres models.
\newline\newline
The evolution of the Szekeres model is given by
\begin{align}
	A_{,t}^2 = \frac{2M}{A} - k +\frac{1}{3c^2}\Lambda A^2
\end{align}
and the density can be written as
\begin{align}
	\rho = \frac{2M_{,r} - 6M\frac{E_{,r}}{E}}{c^2\beta A^2\left(A_{,r} - A\frac{E_{,r}}{E} \right) },
\end{align}
where $\beta := 8\pi G$ and $M(r)$ appears as a constant of integration with respect to time.
\newline\newline
For the specific LTB model considered here, the curvature function $k(r)$ is chosen to have the form 
\begin{equation}\label{eq:k}
k(r) = \left\{ \begin{array}{rl}
-5.4\cdot 10^{-8}r^2\left(\left(\frac{r}{r_b} \right)^6 -1 \right)^6  &\text{if} \,\, \leq r_b:=40\rm Mpc \\
0 &\mbox{ otherwise}
\end{array} \right.
\end{equation}
Since the goal here is to consider a Swiss-cheese model, the LTB model should reduce exactly to a background FLRW model at some $r = r_b$. For this to be possible, $A$  must reduce to $A = ar$ at $r = r_b$, where $a$ is the scale factor of the background. This can e.g. be achieved by, in addition to choosing $k(r)$ as above, choosing $A(t_i,r) = ar$ with $t_i$ the time where the background scale factor is equal to $1/1200$. The considered model includes a cosmological constant with $\Omega_{\Lambda,0} = 0.7$ (setting $H_0 = 70\rm km/s/Mpc$) but this contribution is negligible at early times. Therefore, following \cite{StrongToWeak}, one can set
\begin{align}
	M(r) = \frac{4\pi G\rho_{\rm bg}(t_i)}{3\left( a(t_i)r\right)^3 }\left(1 + \frac{3}{5}\frac{k(r)}{\left(ra(t_i)H(t_i) \right)^2 } \right),
\end{align}
corresponding to the constant big bang time of $t_{\rm bb} = 0$ (i.e. $A(0,r) = 0$), and where $\rho_{\rm bg}$ is the background density which is chosen according to $\Omega_{m,0} = 0.3$ so that the background is flat.
\newline\newline
Although it is mainly the LTB model which will be considered, results based on a Szekeres model will also be presented. This model is obtained by modifying the LTB model by introducing
\begin{align}
	P(r) = Q(r) = 
	\left\{ \begin{array}{rl}
	-1.768\cdot 10^{-5}r^2\left( \left( \frac{r}{r_b}\right)^{10} -1\right)^4   &\text{if} \,\, r\leq r_b \\
	0 &\mbox{ otherwise}
	\end{array} \right.
\end{align}
while setting  $S = 1$.

\subsection{Light propagation}
The redshift along a light ray is given by
\begin{align}
	1+z = \frac{\left(u^{\alpha}k_{\alpha} \right) _e}{\left(u^{\beta}k_{\beta} \right) _o},
\end{align}
where the subscripts $e,o$ indicate the spacetime event of emission and observation of the light ray, respectively. Assuming a non-accelerating velocity field $u^{\alpha}$, the redshift may also be computed as an integration over the local expansion rate and projected shear along the light path. Specifically, the redshift can be computed as 
\begin{align}\label{eq:z_integral}
	1+z = \exp\left(\int_{t_e}^{t_o}dt \left( \frac{1}{3}\theta + \sigma^{\alpha}_{\beta}e_{\alpha}e^{\beta} \right)  \right), 
\end{align}
where, for the quasi-spherical Szekeres model, (a subscripted semi-colon indicates covariant derivative) 
\begin{align}
	\theta := u^{\alpha}_{;\alpha} = \frac{A_{,tr} - 3A_{,t}\frac{E_{,r}}{E} +2\frac{A_{,t}A_{,r}}{A}}{A_{,r}-A\frac{E_{,r}}{E}}
\end{align}
and the non-vanishing components of the shear tensor ($\sigma_{\alpha\beta} := u_{\left(\alpha;\beta \right) } - \frac{1}{3}h_{\alpha\beta}\theta$ with $h_{\alpha\beta} = g_{\alpha\beta} + u_{\alpha}u_{\beta}$) are
\begin{align}
	\sigma^{r}_{r} &= \frac{2}{3}\frac{A_{,tr} - \frac{A_{,t}A_{,r}}{A}}{A_{,r} - A\frac{E_{,r}}{E}}\\
	\sigma^p_p &= \sigma^q_q = -\frac{1}{2}\sigma^r_r.
\end{align}
The vector $e^{\alpha}$ is proportional to the spatial direction of $k^{\alpha}$ as seen by the observer and is given by $e^{\alpha} = \frac{k^{\alpha}}{-u^{\beta}k_{\beta}}-u^{\alpha}$.
\newline\newline
In order to evaluate either of the two expressions for the redshift given above, the light path of the ray/beam must be known. This can be computed by solving the null-geodesic equations  $\frac{d}{d\lambda}\left(g_{\alpha\beta}k^{\beta} \right) = \frac{1}{2}g_{\beta\gamma,\alpha}k^{\beta}k^{\gamma} $, where $\lambda$ is an affine parameter along the light ray.
\newline\indent
The angular diameter distance along the light beam is also needed. This can be computed by introducing the tidal matrix
\begin{equation}
T_{ab} = 
\begin{pmatrix} \mathbf{R}- Re(\mathbf{F}) & Im(\mathbf{F}) \\ Im(\mathbf{F}) & \mathbf{R}+ Re(\mathbf{F})  \end{pmatrix} 
\end{equation}
and solving the transport equation \cite{arb_spacetime}
\begin{equation}
\frac{d^2D^a_b}{d\lambda^2} = T^a_cD^c_b
\end{equation}
simultaneously with solving the null geodesic equations and the parallel propagation equations for the orthonormal vectors spanning screen space, $E^{\mu}_1$ and $E^{\mu}_2$. The latter are needed because they enter into the expressions for the components of the tidal matrix which are given by
\begin{align}
\begin{split}
\mathbf{R}&: = -\frac{1}{2}R_{\mu\nu}k^{\mu}k^{\nu}\\
\mathbf{F}&:=-\frac{1}{2}R_{\alpha\beta\mu\nu}(\epsilon^*)^{\alpha}k^{\beta}(\epsilon^*)^{\mu}k^{\nu},
\end{split}
\end{align}
where $R_{\mu\nu}$ and $R_{\alpha\beta\mu\nu}$ are the Ricci and Riemann tensors, respectively, and $\epsilon^{\mu} := E^{\mu}_1 + iE^{\mu}_2$. The angular diameter distance can then be computed as $D_A= \sqrt{|\det D|}$ if initial conditions are set according to $k^t_i=-1$. The screen-space basis vectors are chosen such that they are orthogonal to each other as well as to $u^{\alpha}$ and $k^{\alpha}$. To fulfill this, their initial conditions are set according to 
\begin{widetext}
\begin{align}
\begin{split}
E^{\mu}_1&\propto \left(0, \frac{\sqrt{F}}{\sqrt{R}}\sqrt{\left(k^p \right) ^2+\left(k^q \right) ^2}, -\frac{\sqrt{R}}{\sqrt{F}}\frac{k^rk^p}{\sqrt{\left(k^p \right) ^2+\left(k^q \right) ^2}}, -\frac{\sqrt{R}}{\sqrt{F}}\frac{k^rk^q}{\sqrt{\left(k^p \right) ^2+\left(k^q \right) ^2}} \right) \\
E^{\mu}_2 & \propto \left(0, 0, \frac{1}{\sqrt{F}}\frac{k^q}{\sqrt{\left(k^p \right) ^2+\left(k^q \right) ^2}} , \frac{-1}{\sqrt{F}}\frac{k^p}{\sqrt{\left(k^p \right) ^2+\left(k^q \right) ^2}}\right) ,
\end{split}
\end{align}
\end{widetext}
where $F:=\frac{A^2}{E^2}$ and $R:=\frac{\left(A_{,r}-A\frac{E_{,r}}{E}\right)^2}{1-k}$ were introduced as short-hand notation for the metric components $g_{pp} = g_{qq}$ and $g_{rr}$.

\begin{figure*}
	\centering
	\subfigure{
	\includegraphics[scale = 0.5]{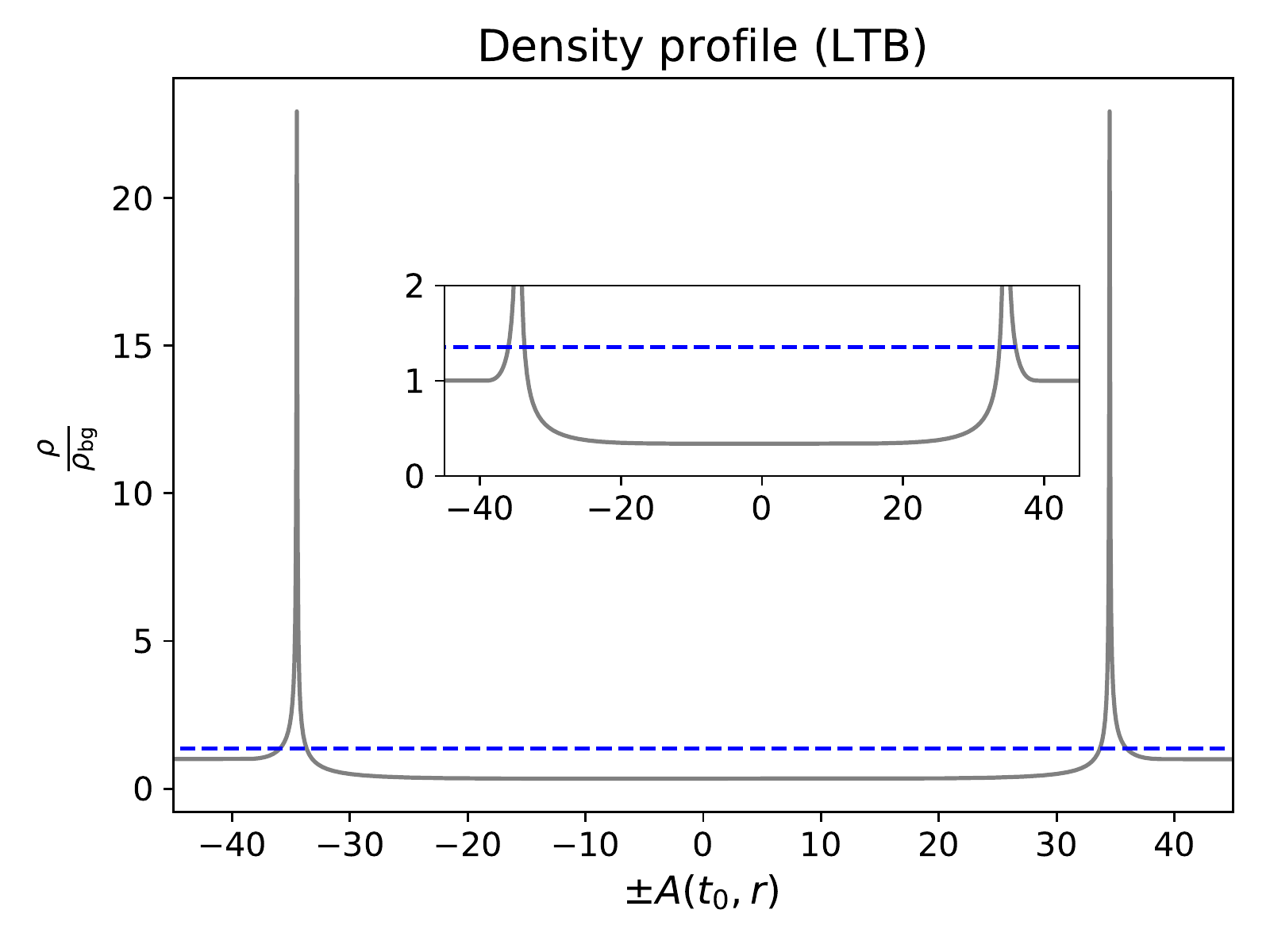}
	}
\subfigure{
	\includegraphics[scale = 0.5]{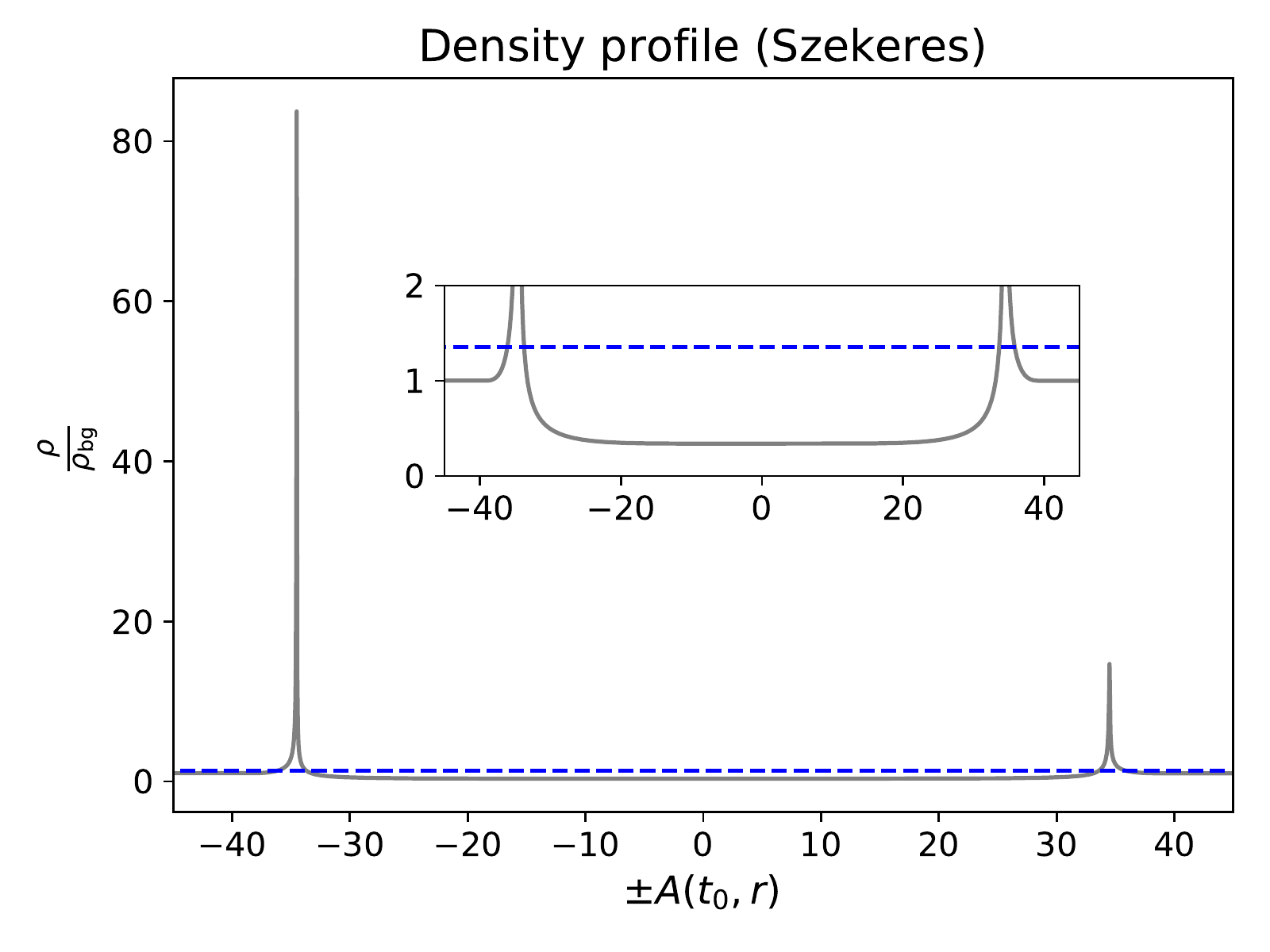}
	}
	\caption{Present time 1D density profiles of LTB and quasi-spherical Szekeres models. For the Szekeres model, the profile is shown along the direction of highest degree of anisotropy. The vertical axis shows $\pm A(t_0,r)$ which approximately corresponds to $\pm$ the physical distance from the origin. The present-time average density field computed without the opaque regions is also shown. The figures include close-ups of the density to more clearly illustrate the central underdensity which has been made opaque to the light rays.}
	\label{fig:profiles}
\end{figure*}

\subsection{Numerical results}
This section serves to present the results obtained by computing the redshift-distance relation along light rays in LTB and Szekeres Swiss-cheese models. The Swiss-cheese models are constructed on-the-fly, with a given light ray initialized at $r = r_0:=45\rm Mpc$ and $t = t_0$ (present time) with an arbitrary impact parameter towards the Szekeres/LTB structure.  Once the light ray again reaches $r = r_0$, it is turned back towards the structure with a new, arbitrary impact parameter and so forth until $z = 2$ is reached. The maximum redshift of $z = 2$ was chosen because it is enough to get a clear picture of the relation between the mean redshift-distance relation, the Dyer-Roeder approximation and the redshift-distance relation based on spatial averages.
\newline\indent
Since the goal is to consider the mean redshift-distance relation in the presence of opaque regions, the central part of the structures are made opaque by requiring that if a light ray gets closer to the center of an inhomogeneous region than $r_{\rm lim}:=25\rm Mpc$, the light ray is moved back to the previous point along the light path where it reached $r = r_0$. The light ray is then propagated towards the structure again with a new (random) impact parameter. Figure \ref{fig:profiles} shows present-time LTB and Szekeres density profiles. Densities are normalized such that a density of 1 corresponds to the density of the background FLRW model. In these units, the spatial average (also shown) is roughly $1.35$ i.e. roughly 35 \% greater than the background density. In the Dyer-Roeder approximation this would correspond to $\alpha(t = t_0)=1.35>1$, in accordance with the central underdense regions being opaque. Note though, that normally $\alpha<1$ is considered for real narrow light beams since these are expected to mostly travel through voids and avoid regions with larger density. This difference has no significant implications for the presented results as it is of no principal significance for the validity of the Dyer-Roeder approximation whether $\alpha$ is greater or less than 1. The important point is that the model contains regions which are opaque to the light rays.
\newline\newline

\begin{figure}
	\centering
		\includegraphics[scale = 0.55]{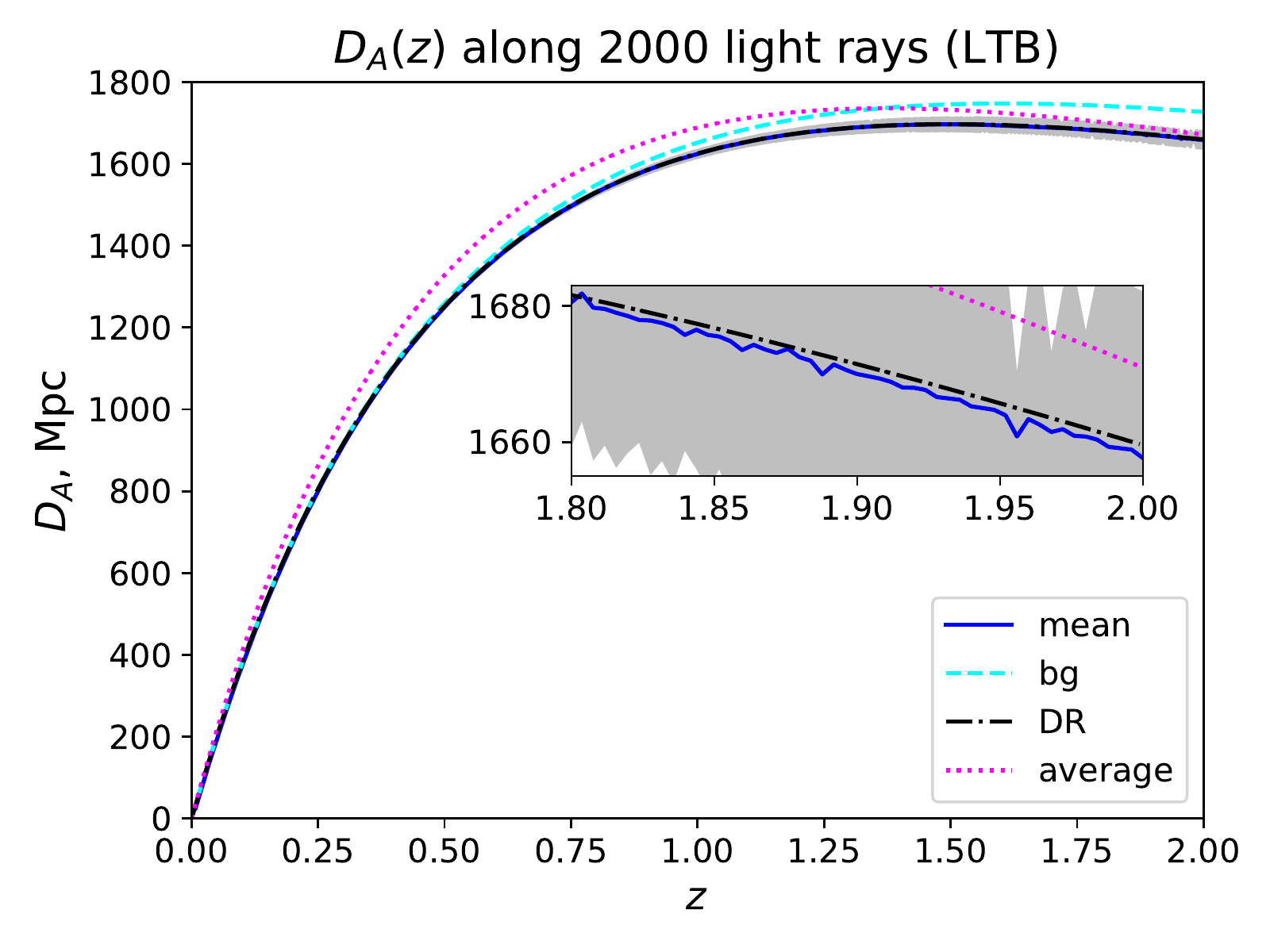}
	\caption{Mean (black line) and spread (gray shaded area) of redshift-distance relation along 2000 light rays with comparison with prediction by the Dyer-Roeder approximation (
	``DR''), the relation based on spatial averages (``average'') and the background (``bg''). The shaded area, the Dyer-Roeder expression and the mean redshift-distance relation are barely distinguishable in the main figure so a close-up is included of the last part of the light ray to illustrate the small difference between the two latter.}
	\label{fig:DAz}
\end{figure}
\begin{figure}
	\centering
	\subfigure{
		\includegraphics[scale = 0.55]{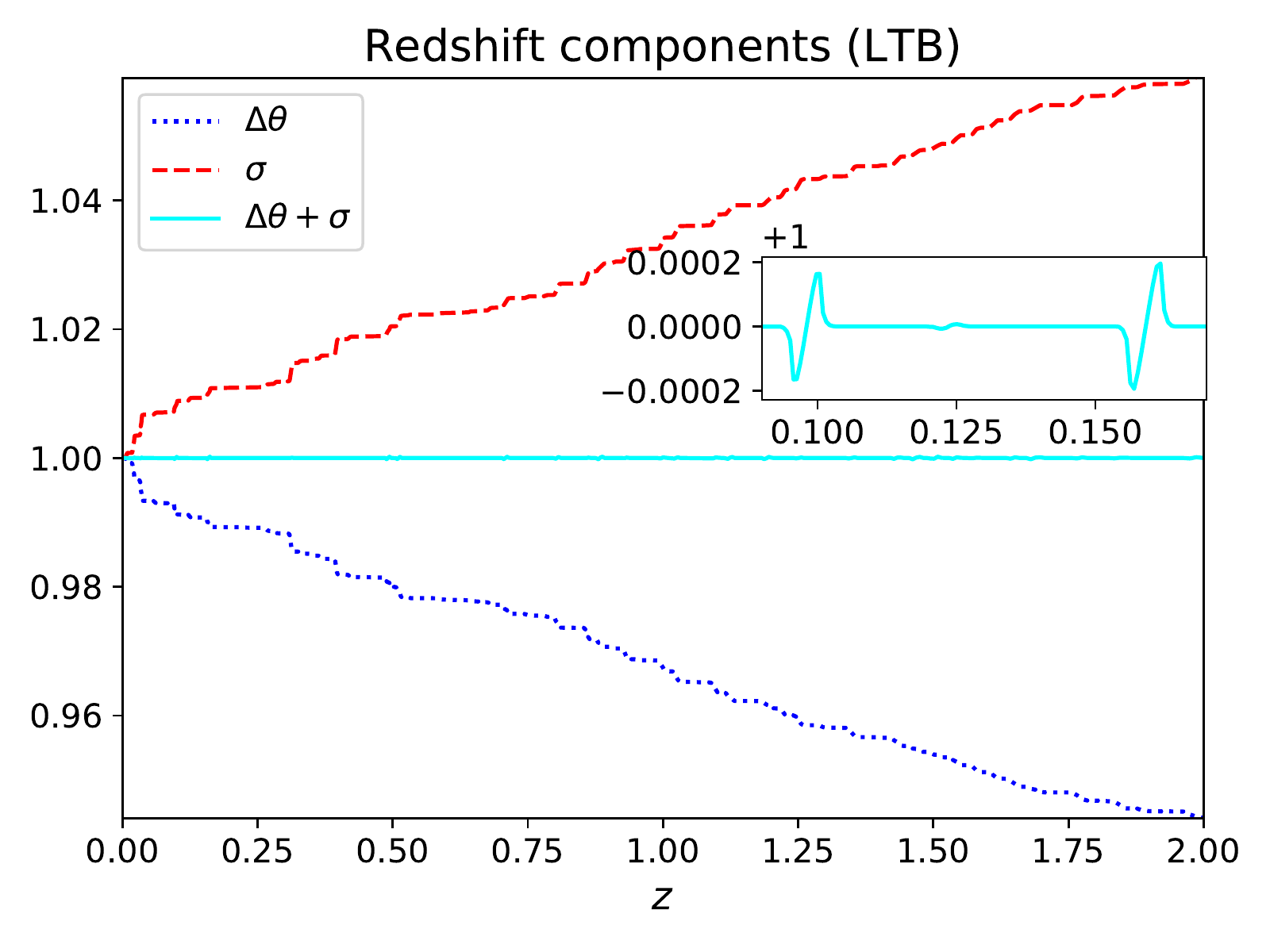}
	}
	\subfigure{
		\includegraphics[scale = 0.55]{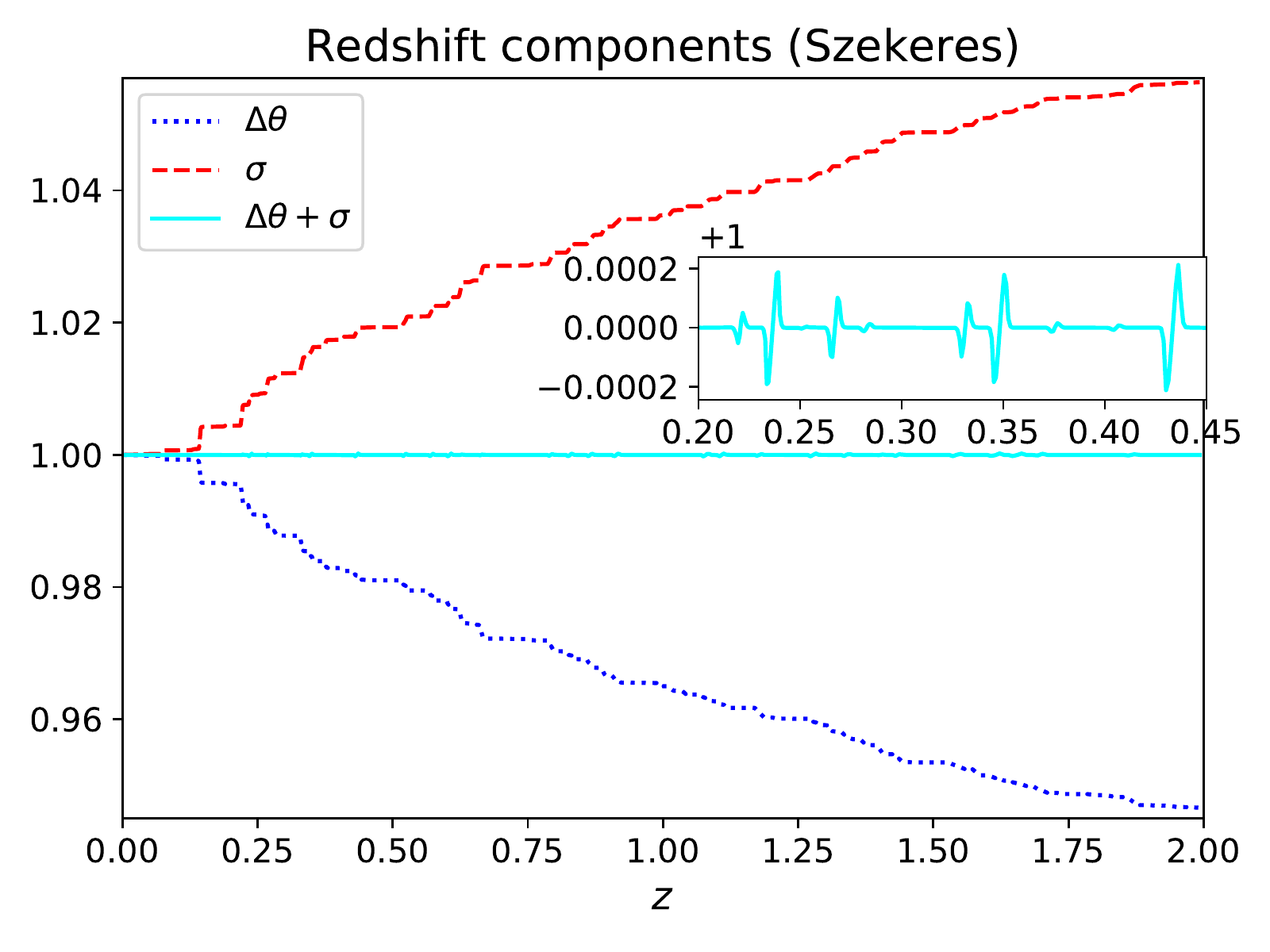}
	}
	\caption{Components of the redshift fluctuation according to $
		\frac{1+z}{1+z_{\rm bg}} = \exp\left(\int_{t_e}^{t_o}dt \left( \frac{1}{3}\Delta \theta + \sigma^{\alpha}_{\beta}e_{\alpha}e^{\beta} \right)  \right)$ for single light rays in the Swiss-cheese models based on LTB and Szekeres structures. The contribution $\exp\left(\int_{t_e}^{t_o}dt \frac{1}{3}\Delta \theta  \right)$ is denoted by $\Delta \theta$  while the contribution $\exp\left(\int_{t_e}^{t_o}dt  \sigma^{\alpha}_{\beta}e_{\alpha}e^{\beta}   \right)$ is denoted $\sigma$ (and $\Delta\theta:=\theta-\theta_{\rm bg}$). The total fluctuation contribution, $\exp\left(\int_{t_e}^{t_o}dt \left( \frac{1}{3}\Delta \theta + \sigma^{\alpha}_{\beta}e_{\alpha}e^{\beta} \right)  \right)$, is denoted by $\Delta\theta+\sigma$. Close-ups are included to show the small deviations of the total fluctuation contribution from 1.
	}
	\label{fig:z_cheese}
\end{figure}

The redshift-distance relation has been computed along 2000 light rays in the Swiss-cheese model with LTB structures. The resulting mean and dispersion of the redshift-distance relation is shown in Figure \ref{fig:DAz} where it is compared with the redshift-distance relation of the background FLRW model and the predictions of the Dyer-Roeder equation (Eq. \ref{eq:DR}) as well as with the predictions of the redshift-distance relation based on spatially averaged quantities (Eq. \ref{eq:av},\ref{eq:syksy_z}). When computing the Dyer-Roeder approximation, the Ricci-term was computed using the spatial average of the density field of the transparent regions. The same is done when computing the redshift-distance relation based on spatial averages but in that case, also the redshift and expansion rates are computed according to spatial averages of the transparent regions. In the Dyer-Roeder approximation, the redshift and expansion rate are computed according to the background FLRW model.
\newline\indent
As seen in Figure \ref{fig:DAz}, the Dyer-Roeder approximation makes a good prediction for the mean redshift-distance relation and is indeed nearly indistinguishable from the mean redshift-distance relation. The same is not true for the relation based on spatial averages which deviates significantly from the actual mean redshift-distance relation. This is exactly as anticipated since it has earlier been noted \cite{tardis, cmb_SMK} that the expansion rate fluctuations compared to the background expansion rate to a high precision cancel with the projected shear contribution to the redshift locally along light rays. This is illustrated in Figure \ref{fig:z_cheese} for a single light ray in the Swiss-cheese models based on the LTB and Szekeres structures. The local deviation from the background is sub-percent everywhere despite the fluctuations in the expansion rate and the projected shear individually contribute to fluctuations of the redshift at percent order. In addition, the small sub-percent deviations along the light rays are only visible while the light ray is actually inside the structure i.e. the fluctuations that accumulate while a light ray traverses a structure cancel to a very high precision upon traversal of an entire structure. This is similar to what was found in \cite{DR_SMK} for Einstein-Straus models except that there, the main part of the fluctuations in the redshift occurred on the boundary between Schwarzschild and FLRW regions.
\newline\indent
Only a single light ray has been considered for the Swiss-cheese model based on Szekeres structures since Figure \ref{fig:z_cheese} illustrates that also for the case of the anisotropic quasi-spherical Szekeres structures does the redshift locally equal the background redshift to a high precision (and the spatial averages of the Szekeres model are equal to those of the corresponding LTB model as discussed in e.g. \cite{av_Szekeres}). Thus, the result shown in Figure \ref{fig:DAz} for the Swiss-cheese model with LTB structures can safely be extrapolated to also be valid for Swiss-cheese models based on (quasi-spherical) Szekeres models.

\section{N-body simulations and perturbation theory}\label{sec:nbody}
The Swiss-cheese models are interesting to study because they are exact solutions to the Einstein equations which can be constructed as inhomogeneous cosmological models with a statistically homogeneous and isotropic distribution of matter. The inhomogeneities described by the models are, however, too simple for the resulting models to be considered particularly realistic. While Swiss-cheese models are therefore useful for especially proof-of-principle studies, it is often desirable to also study more realistic structure formation scenarios.
\newline\indent
N-body simulations are generally considered to yield the most realistic renderings of our universe as they can be initialized based on random fluctuations on an FLRW background at early times (in accordance with CMB observations) and propagated forwards to evolve into a network of non-linear structures at late times. Originally, these numerical simulations were Newtonian (e.g. \cite{ramses,gadget2,concept}) but more recently also relativistic codes for studying cosmic evolution have appeared \cite{gevolution,rel1,rel2,rel3,rel4,rel5,rel6,rel7}.
\newline\newline
The goal with this section is to confirm that the Dyer-Roeder approximation is also valid within the setting of more complicated structures generated with numerical N-body codes. Since neither relativistic species, extremely high precision details including vector and tensor modes nor non-standard cosmological metrics are of interest here, it is sufficient to use a Newtonian N-body simulation for the current work (see e.g. \cite{gevolution,relNewt_compare,gevolution_nature} for comparisons of Newtonian and general relativistic codes). The numerical results presented in this section were therefore obtained with Gadget-2 \cite{gadget2,gadget}.
\newline\indent
The following subsections serve to describe the considered N-body simulation including how it is related to light propagation, as well as presenting numerical results.

\subsection{The N-body simulation and light propagation formalism}
To study the mean redshift-distance relation along light rays in an N-body simulation, Gadget-2 was run with an Einstein-de Sitter (EdS) background using initial conditions generated with N-GenIC\footnote{The N-GenIC code can be downloaded from https://wwwmpa.mpa-garching.mpg.de/gadget/.}. An EdS background was used in order to enhance the effects of inhomogeneities which are naturally suppressed if, say, $70\%$ of the present-day density is in the form of a homogeneously distributed dark energy component. The code was run using $512^3$ particles on a $512{\rm Mpc}/h$ grid (with $h$ the reduced Hubble parameter, set to 0.7). 24 snapshots corresponding roughly to a background redshift interval of $0\leq z\leq1$ were obtained and saved. The triangular shaped cloud (TSC) method was used to interpolate from the discrete masses to a density field and was also used to construct a corresponding smooth velocity field.
\newline\newline
Since the goal with the snapshots is to describe light propagation, a relation between the Newtonian snapshots and a corresponding relativistic spacetime is required. The connection between Newtonian N-body simulations and corresponding relativistic counterparts has been studied in e.g. \cite{NewtVsRel1,NewtVsRel2,NewtVsRel3,NewtVsRel4,NewtVsRel5,NewtVsRel6}. Again since extremely high precision (and super-horizon scales \cite{redshift_Nbodygauge}) are not of interest here, it is sufficient to consider the simple recipe of \cite{NewtVsRel2}. In this case, the spacetimes corresponding to the snapshots are identified with the perturbed FLRW metric in the Newtonian gauge which we can write as
\begin{align}
	ds^2 = -c^2(1+2\psi)dt^2 + a^2(1-2\psi)\left(dx^2 + dy^2 + dz^2 \right),
\end{align}
where $\psi$ is computed according to $\nabla^2\psi = 4\pi G a^2 \delta\rho$, with $\delta\rho$ the overdensity of the TSC-generated density field compared to the EdS background. To obtain $\psi$ in practice, this equation is solved in Fourier space using FFTW3\footnote{http://www.fftw.org/}.
\newline\indent
With the above procedure, the density field, velocity fields and $\psi$ are computed on a 4-dimensional grid. Their values at any point along a light ray are obtained by quadri-linear interpolation.
\newline\newline
4600 light rays were traced through the N-body simulation (using periodic boundary conditions) with each light ray corresponding to a randomly placed observer (always placed at $t = t_0$) looking in a random direction. Light rays were then propagated by using the full null-geodesic equations corresponding to the perturbed metric described above. The corresponding redshift along the light rays can be computed as $1+z = \frac{\left(u^{\alpha}k_{\alpha} \right) _e}{\left(u^{\beta}k_{\beta} \right) _o}$, with $u^{\alpha}\propto \left(1, v^{i} \right) $, where $v^i$ is the velocity field obtained with the TSC method and where $u^{\alpha}$ is normalized according to $u^{\alpha}u_{\alpha}=-1$.
\newline\indent
It is well-known that the mean redshift-distance relation of light rays propagated through N-body simulations corresponds well with the background relation if all regions of the snapshots are treated as transparent (see e.g. \cite{Fleury_Nbody} and references therein). The issue studied here is what happens if some regions are made opaque. It is not currently feasible to study an N-body simulation with a resolution high enough to actually mimic the distribution of matter along narrow light beams so opaque regions are introduced more practically by simply not permitting light rays to travel in regions with $\delta:=\frac{\delta\rho}{\rho_{\rm EdS}}\geq0$. In practice, this is implemented by turning a light ray around if it reaches a region with $\delta\geq0$. This is clearly not fantastically realistic but it is a fairly fast and straight forward way to obtain a large number of light rays that sample the desired spacetime. As long as a large number og light rays is considered, the non-realistic features of the method should not affect the mean although it will most likely exaggerate the spread in the redshift-distance relation since some light rays will be ``trapped'' in a relatively small region between structures with $\delta\geq0$.
\newline\newline
As with the Swiss-cheese models, the angular diameter distance is computed by solving the transport equation along the light rays. To speed up the computations, the Weyl term ($\mathbf{F}$) in the tidal matrix is neglected. With as small number of light rays it has been confirmed that the Weyl lensing has negligible effect on the results.

\begin{figure}
	\centering
	\includegraphics[scale = 0.55]{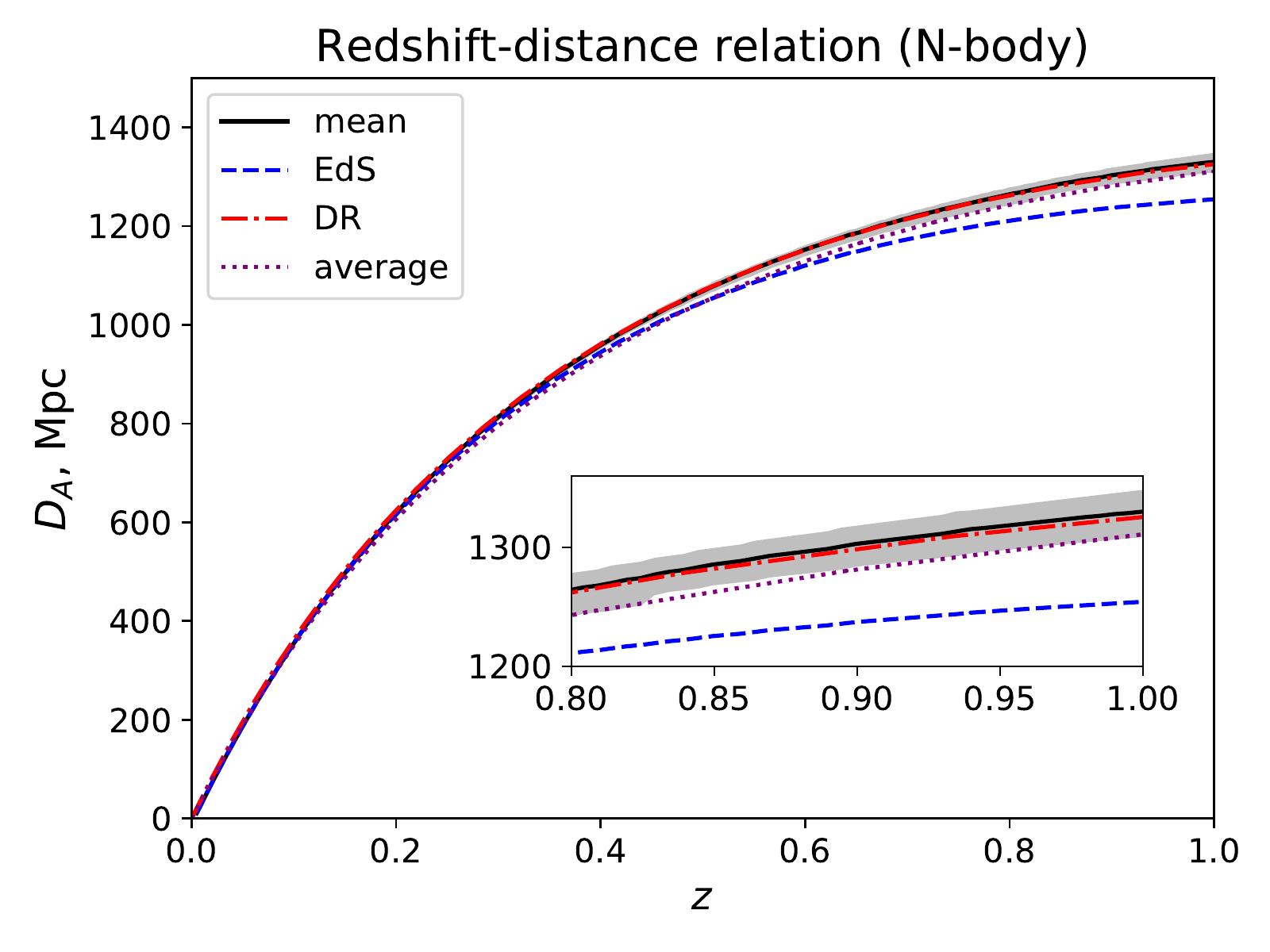}
	\caption{Mean (black line) and spread (gray shaded area) of redshift-distance relation along light rays propagated through N-body simulation. The figure also shows the redshift-distance relation of the EdS model (the background of the N-body simulation), as well as the predictions based on the Dyer-Roeder (``DR'') approximation and the spatial averages of the transparent regions (``average''). A close-up is shown since is is difficult to distinguish between several of the lines in the main figure.}
	\label{fig:DAz_nbody}
\end{figure}

\subsection{Numerical results}
\begin{figure}
	\centering
	\includegraphics[scale = 0.55]{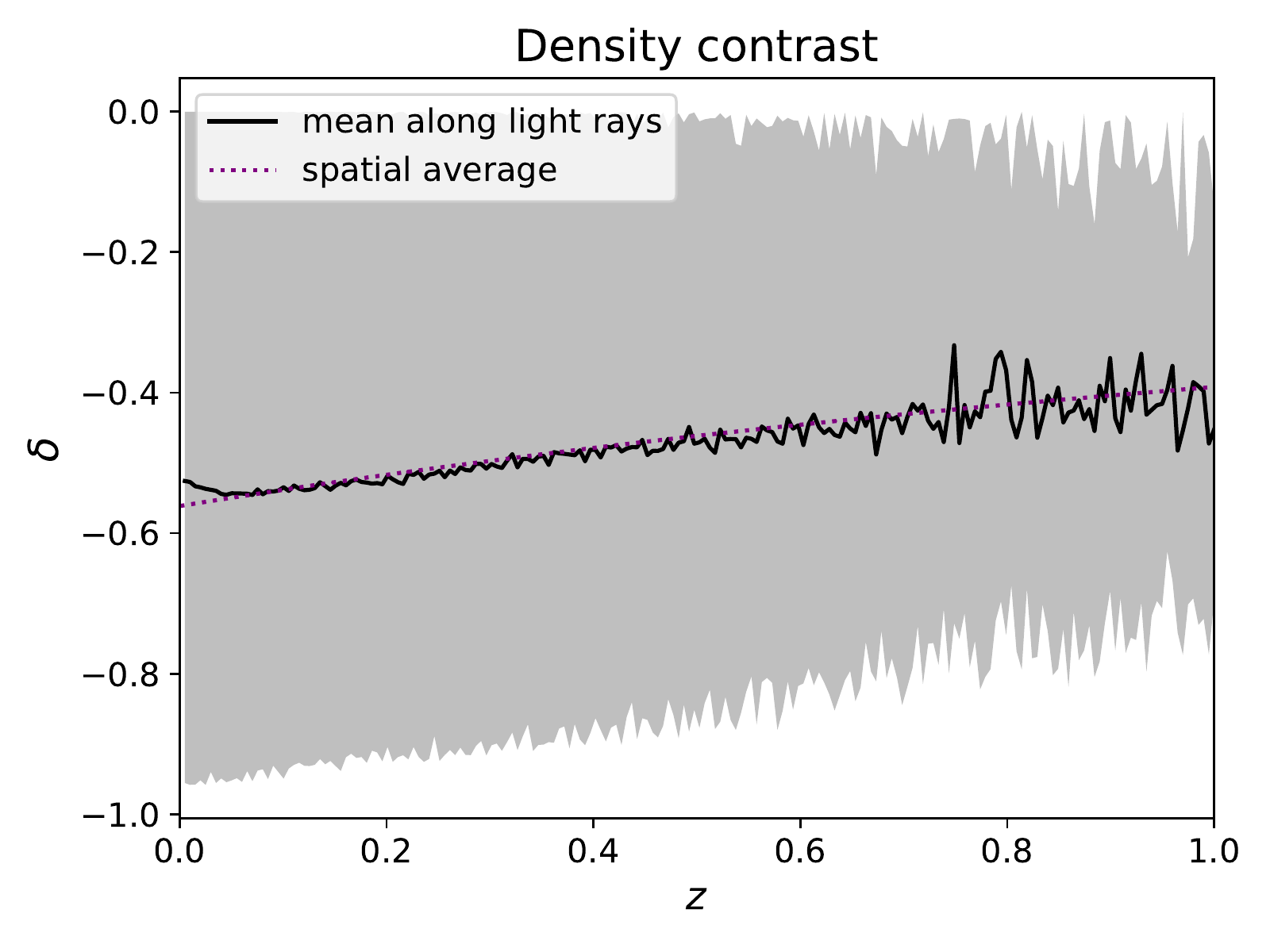}
	\caption{Mean and spread of density along light rays in N-body simulation compared with the spatial average of the density of the transparent regions. The dotted line representing the latter is difficult to see since the mean density fluctuates very closely around the spatial average of the density of the transparent region.}
	\label{fig:density_nbody}
\end{figure}

Figure \ref{fig:DAz_nbody} shows the mean and spread of the redshift-distance relation obtained along the light rays propagated through the cosmological model described by the snapshots. For comparison, the redshift-distance relation of the EdS model as well as the predictions of the Dyer-Roeder approximation and that of the method based on spatial averages (of the transparent regions) are included. There is a good (but not perfect) agreement between the mean relation and the prediction of the Dyer-Roeder approximation. The prediction based on the spatial averages is also fairly good, although not nearly as good as the prediction based on the Dyer-Roeder approximation.
\newline\indent
\begin{figure}
	\centering
	\includegraphics[scale = 0.55]{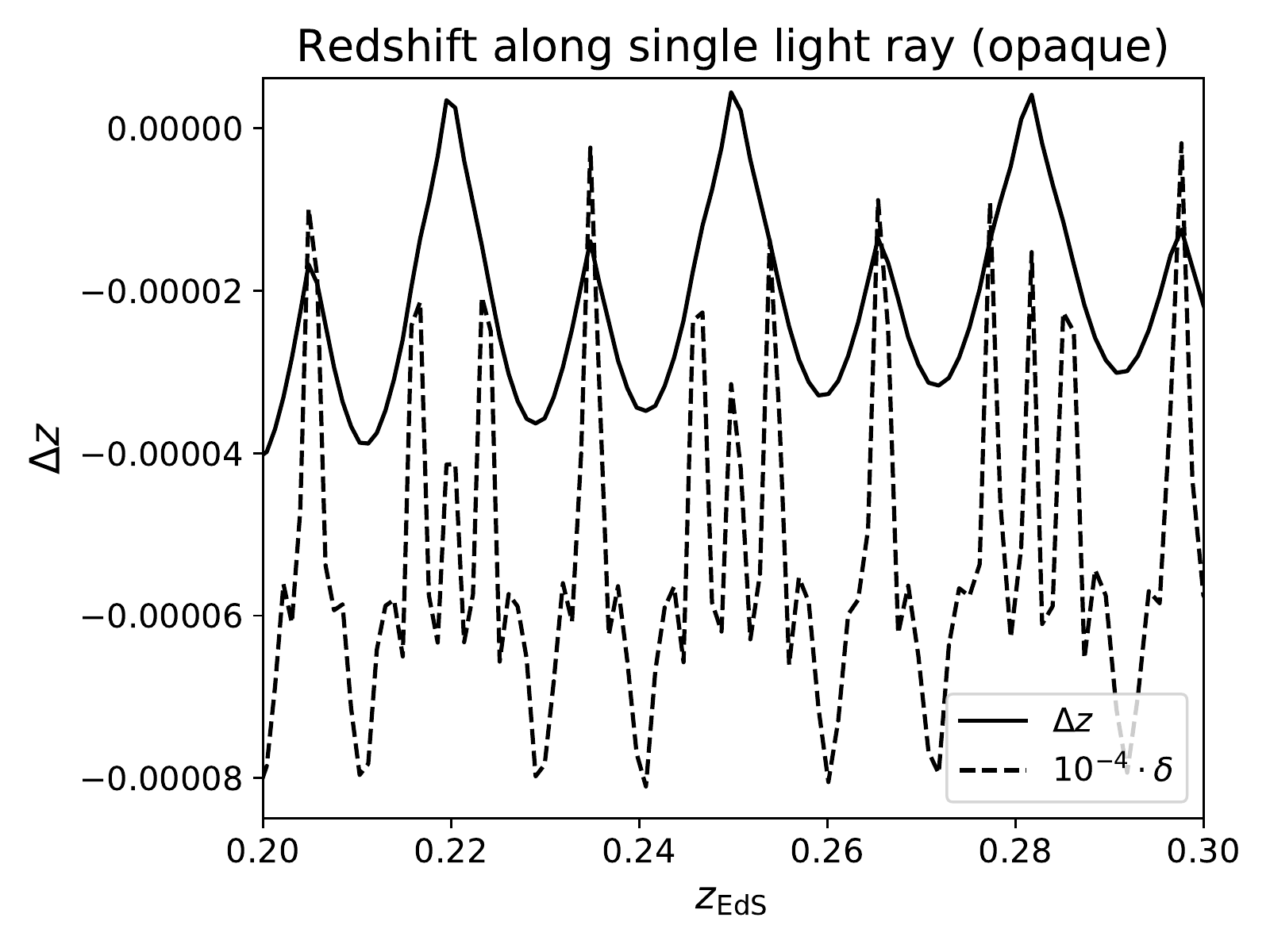}
	\caption{The redshift along a single light ray in the N-body simulation with opaque regions. The redshift is shown as the fluctuation about the EdS redshift, i.e. $\Delta z:=\frac{z-z_{\rm EdS}}{z_{\rm EdS}}$. The redshift fluctuation is shown together with a scaling of the density contrast.}
	\label{fig:redshift_nbody}
\end{figure}
Just as for the Swiss-cheese models, the good agreement between the Dyer-Roeder approximation and the mean redshift-distance relation can be understood by looking at the redshift. First of all, one may note that the studied spacetime is near-FLRW as defined in \cite{syksy_nearFRW}. For such spacetimes, it was shown in \cite{syksy_nearFRW} that the dominant part of the fluctuation in the expansion rate cancels with the dominant contribution from the projected shear (see specifically Eq. 5.8 and the related discussion). Indeed, as shown in Figure \ref{fig:redshift_nbody} the redshift along an individual light ray is everywhere very close to the background redshift. Specifically, the figure shows $\Delta z:=\frac{z-z_{\rm EdS}}{z_{\rm EdS}}$ and the corresponding density field along a small portion of a fiducial light ray. Just as was seen for the Swiss-cheese models, the redshift fluctuates slightly, with the fluctuations due to a light ray moving into a structure compensated/canceled by the fluctuations due to the redshift moving out of the structure. In addition one may note that the fluctuations do not appear to be around exactly zero. This is not surprising as the redshift must have a small offset due to the Reese-Sciama/Integrated Sachs-Wolf effect \cite{ISW1,ReeseSciama,ISW_review}, implying that the redshift cannot be {\em exactly} equal to the background redshift. 
\newline\indent
The small difference between the Dyer-Roeder approximation and the actual mean redshift-distance relation is presumably partially due to the fairly low precision used for the computations. It can also partially be understood as a consequence of the mean density along the computed light rays deviating slightly from the spatial mean. This deviation is shown in Figure \ref{fig:density_nbody} and could presumably be avoided by including more light rays.
\newline\indent
The spatial average of the expansion rate
\footnote{The spatial average of the expansion rate is approximated as  $\frac{\sum \left( 1-3\psi\right) \left( 3H+v^i_{,i}\right) }{\sum 1-3\psi}$ with the sum including grid points with $\delta <0$. This approximation takes only the dominant part of the local fluctuation in the expansion rate into account but it has been verified that taking subdominant contributions into account leads to no visible difference. Note also that the averaging procedure as well as the computation of the redshift using Eq. \ref{eq:syksy_z} is slightly incorrect since it ignores the deviation of the lapse function from 1. The consequences of this are expected to be negligible. Note that the same approximation of spatial averages was used when computing the spatially averaged density used for both the Dyer-Roeder approximation and the relation based on spatial averages.}
deviates slightly from the background expansion rate. This then means that the redshift based on spatial averages is slightly different than that of the background. This explains why the redshift-distance relation based on spatial averages does a (slightly) poorer job than the Dyer-Roeder approximation in describing the mean redshift-distance relation.
\newline\newline
It must be noted that a higher resolution than that used here could in principle lead to a more significant deviation between the Dyer-Roeder approximation and the mean redshift-distance relation. This would especially be possible if density contrasts large enough to lead to a significant contribution from Weyl lensing were included. However, the main point with this section is to illustrate that the redshift along light rays in this type of spacetime are well approximated by the background redshift even when opaque regions are introduced. It seems quite unlikely that a higher resolution would significantly alter this point as it is understood through the analytical arguments in \cite{syksy_nearFRW} to be valid largely as long as the metric remains perturbatively close to the background FLRW metric which is expected to be realistic as long as one does not consider highly dense structures such as neutron stars or black holes. On the other hand, estimating if Weyl lensing becomes important for some real observations requires actually being able to resolve structures on the appropriate (very small) scale.

\section{A model without local cancellation of projected shear and expansion rate fluctuations}\label{sec:hellaby}
The previous two sections showed that the Dyer-Roeder approximation agrees well with the mean redshift-distance relation of standard inhomogeneous cosmological models, and it was argued that the reason the redshift is well described by that of the background even in models with opaque regions is that the shear contribution to the redshift locally cancels almost exactly with the fluctuations in the expansion rate along individual light rays. Since this conclusion is based on several types of Swiss-cheese models, perturbations theory and N-body simulations, it seems fair to conclude that the Dyer-Roeder approximation can be expected to be valid in standard cosmological scenarios. This section explores the validity of the Dyer-Roeder approximation in the less-standard scenario of an inhomogeneous cosmological model which does not have an FLRW background and where the shear and fluctuations in the expansion rate do not cancel locally along individual light rays. Specifically, the model presented in \cite{Hellaby} is studied. This model was first studied in terms of its light propagation qualities in \cite{Hellaby_light} where it was found that the mean redshift-distance relation is well described by spatially averaged quantities according to Eq. \ref{eq:av}, \ref{eq:syksy_z}. However, the study did not consider the possibility of opaque regions. This is remedied here.
\newline\newline
In the following subsections, the studied model is described together with a formalism for computing the redshift-distance relation in the model. In the final subsection, numerical results are presented and discussed.

\subsection{Model construction}
\begin{figure*}
	\centering
		\includegraphics[scale = 0.55]{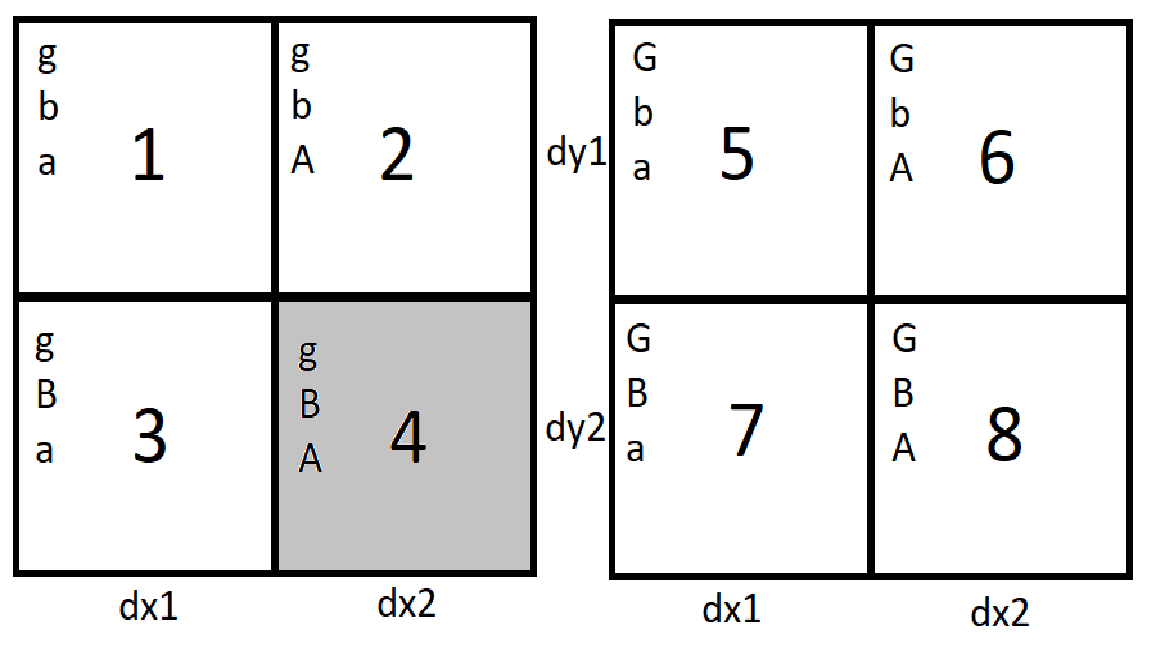}

	\caption{(Modification of figure from \cite{Hellaby_light}.) 2D rendering of fundamental block with values of $\alpha, \beta$ and $\gamma$ indicated by $a$, $A$ $b$, $B$, $g$ and $G$. The cubes on the left side of the figure have a comoving height of dz1 while those to the left have a comoving height of dz2. Values of $a, A$ etc. are given in Table \ref{table:hellaby_model}. The two sets of four cubes are stacked on top of each other such that cube 1 is below cube 5 etc.. The gray-scaled cube, cube 4, is treated as opaque during light propagation.
	}
	\label{fig:hellaby_model}
\end{figure*}

\begin{table*}
	\caption{Model parameters of cubes in the fundamental block of the studied model. The parameters a, A, b, B, g and G refer to values of $\alpha$, $\beta$ and $\gamma$ in different cubes while dx1, dx2 etc. refer to comoving side lengths of the different cubes. See Figure \ref{fig:hellaby_model} for an illustration. }
	\centering
	\begin{tabular}{c c c c c}
	\hline\hline
	(a,b,g) & (A,B,G) & (dx1,dy1,dz1) & (dx2,dy2,dz2)& $t_0$ (Gyr) \\
	\hline
	($1,1,1$) & $-0.05\cdot(1,1,-1)$ &  $30\cdot$(1,1,1) &$10\cdot$(1,1,1) &  8.9 \\
	\hline
\end{tabular}
	\label{table:hellaby_model}
\end{table*}

The model considered in this section is constructed by tessellating space by a simple type of Bianchi I models, each with a line element given by
\begin{align}
ds^2 = -dt^2 + \left(\frac{t}{t_0} \right) ^{2\alpha}dx^2 + \left(\frac{t}{t_0} \right) ^{2\beta}dy^2  + \left(\frac{t}{t_0} \right) ^{2\gamma}dz^2.
\end{align}
Present time is denoted by $t_0$ and $\alpha, \beta$ and $\gamma$ are constants. The line element corresponds to a homogeneous but anisotropic spacetime containing a comoving perfect fluid with anisotropic pressure. Specifically, the density is given by
\begin{align}
	8\pi G\rho =\frac{\alpha\beta + \beta\gamma + \alpha\gamma}{t^2} 
\end{align}
while the pressure components are
\begin{align}
\begin{split}
8\pi G p_x & = \frac{\beta + \gamma - \left(\beta^2 + \beta\gamma + \gamma^2 \right) }{t^2}\\
8\pi G p_y & = \frac{\gamma + \alpha - \left( \gamma^2 + \gamma\alpha + \alpha^2 \right) }{t^2}\\
8\pi G p_z &= \frac{\alpha + \beta - \left( \alpha^2 + \alpha\beta + \beta^2 \right) }{t^2}.
\end{split}
\end{align}
The local expansion rate is given by
\begin{align}
	\theta = \frac{\alpha+\beta+\gamma}{t}
\end{align}
and the non-vanishing components of the shear tensor are
\begin{align}
\begin{split}
\sigma^x_x&= \frac{1}{3t}\left(2\alpha - \beta-\gamma \right) \\
\sigma^y_y &= \frac{1}{3t}\left( 2\beta - \alpha-\gamma \right) \\
\sigma^z_z &=\frac{1}{3t}\left(2\gamma - \alpha-\beta \right) .
\end{split}
\end{align}
As discussed in \cite{Hellaby}, finite spatial cubes can be arranged into a ``fundamental block'' which can be repeated to tessellate the entire spacetime. To fulfill the Darmois junction condition, the metric parameters $\alpha, \beta$ and $\gamma$ corresponding to a direction orthogonal to a junction between two cubes must be constant across the junction. This can e.g. be achieved by constructing a fundamental block of eight cubes arranged according to Figure \ref{fig:hellaby_model}. One specific model is studied here, with model parameters given in Table \ref{table:hellaby_model}. Note that the metric functions are normalized to present time so that the values of dx1, dy1 etc. in Table \ref{table:hellaby_model} give the proper dimensions of the different regions at present time (in Mpc). Thus, the proper side lengths of the individual regions are either 10Mpc or 30Mpc at present time. Going back in time, some of these regions will have side lengths that grow (slowly) but most side lengths will decrease backwards in time. This model was also studied in \cite{Hellaby_light} where it was found that the redshift-distance relation was well described by Eq. \ref{eq:av}, \ref{eq:syksy_z}, with Eq. \ref{eq:av} modified to include the pressure components according to
\begin{align}
\begin{split}
\frac{d^2D_A}{dz_{\rm av}^2} + \left(\frac{2}{1+z_{\rm av}}+\frac{d\ln  H_{\rm av} }{dz_{\rm av}} \right) \frac{dD_A}{dz_{\rm av}} \\ = -\frac{4\pi G\left( \rho_{\rm av} + p_{\rm av} \right) }{H_{\rm av}^2(1+z_{\rm av})^2}D_A,
\end{split}
\end{align}
where the appropriate average of the anisotropic pressure is set as $p_{\rm av} := \frac{1}{3}\left(  \left( p_x\right)_{\rm av}  + \left( p_y\right) _{\rm av} + \left( p_z\right) _{\rm av} \right)  $. When including opaque regions, the pressure is treated similarly to the density, i.e. the mean pressure is computed only using the transparent regions - both when considering the expression based on spatial averages and the Dyer-Roeder approximation.   

\subsection{Light propagation formalism and initial numerical considerations}
\begin{figure}
	\centering
	\includegraphics[scale = 0.5]{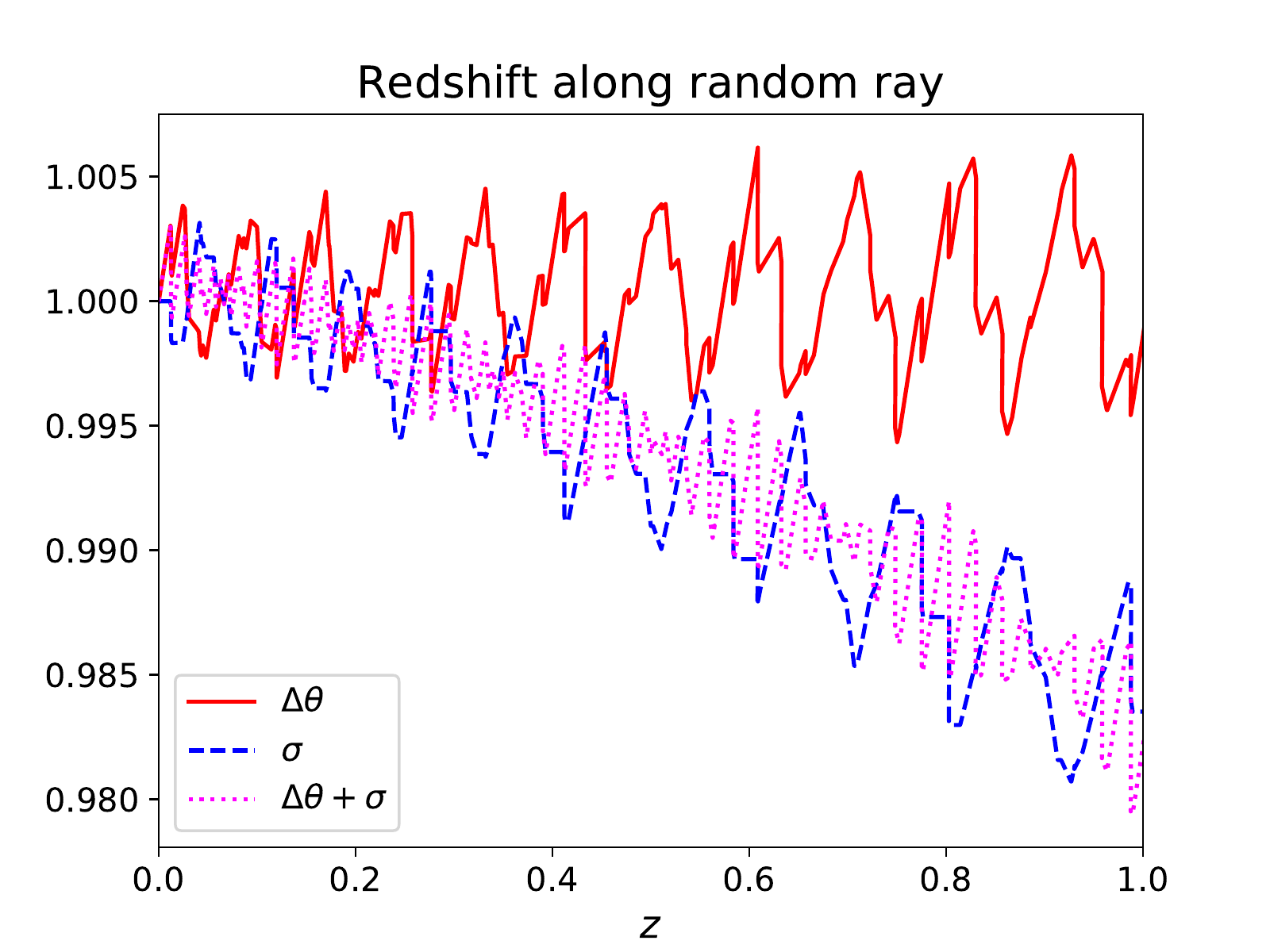}
	\caption{Components of the redshift along a single light ray according to the splitting $
		\frac{1+z}{1+z_{\rm opaque}} = \exp\left(\int_{t_e}^{t_o}dt \left( \frac{1}{3}\Delta \theta + \sigma^{\alpha}_{\beta}e_{\alpha}e^{\beta} \right)  \right)$, where $z_{\rm opaque}$ is the redshift computed from the spatial average of the expansion rate, only including transparent regions. The contribution $\exp\left(\int_{t_e}^{t_o}dt \frac{1}{3}\Delta \theta  \right)$ is denoted by $\Delta \theta$  while the contribution $\exp\left(\int_{t_e}^{t_o}dt  \sigma^{\alpha}_{\beta}e_{\alpha}e^{\beta}   \right)$ is denoted $\sigma$. The total fluctuation contribution, $\exp\left(\int_{t_e}^{t_o}dt \left( \frac{1}{3}\Delta \theta + \sigma^{\alpha}_{\beta}e_{\alpha}e^{\beta} \right)  \right)$, is denoted by $\Delta\theta+\sigma$.}
	\label{fig:redshift_hellaby}
\end{figure}
The light propagation formalism used to compute the redshift-distance relation for the model follows the formalism used in the previous sections, i.e. the light paths are obtained by solving the null-geodesic equations and the angular diameter distance is obtained through solutions of the transport equation including parallel propagation of the orthogonal screen-space basis vectors. One noteworthy point is that the null-geodesic equations (and hence also the parallel propagation equations) contain delta-function contributions. For instance, if a boundary exists at some value $x_b$ of $x$ then the geodesic equation for $k^x$ will contain a term of the form
\begin{align}
	-\frac{1}{2}\frac{g_{xx,x}}{g_{xx}} k^x = -\left( k^x\right) ^2\log\left( t/t_0\right)\delta(x-x_b)\Delta \alpha, 
\end{align}
where $\Delta\alpha$ is the change in the metric parameter $\alpha$ across the boundary at $x = x_b$. By integration, it is seen that this leads to a jump in $k^x$ on the boundary given by
\begin{align}
	k^x \rightarrow k^x+k^x\log\left(t/t_0 \right) \Delta\alpha.
\end{align}
Similar delta-function contributions will of course occur on boundaries at constant $y$ and $z$. As discussed in \cite{Hellaby_light} it is numerically seen that the effect these delta-function contributions have is to re-normalize the tangent (and screen space) vectors so that they stay null (normal) when crossing junctions. There are no delta-function contributions to the Riemann tensor and hence no delta-functions appear in the transport equation.
\newline\newline
\begin{figure}
	\centering
	\includegraphics[scale = 0.5]{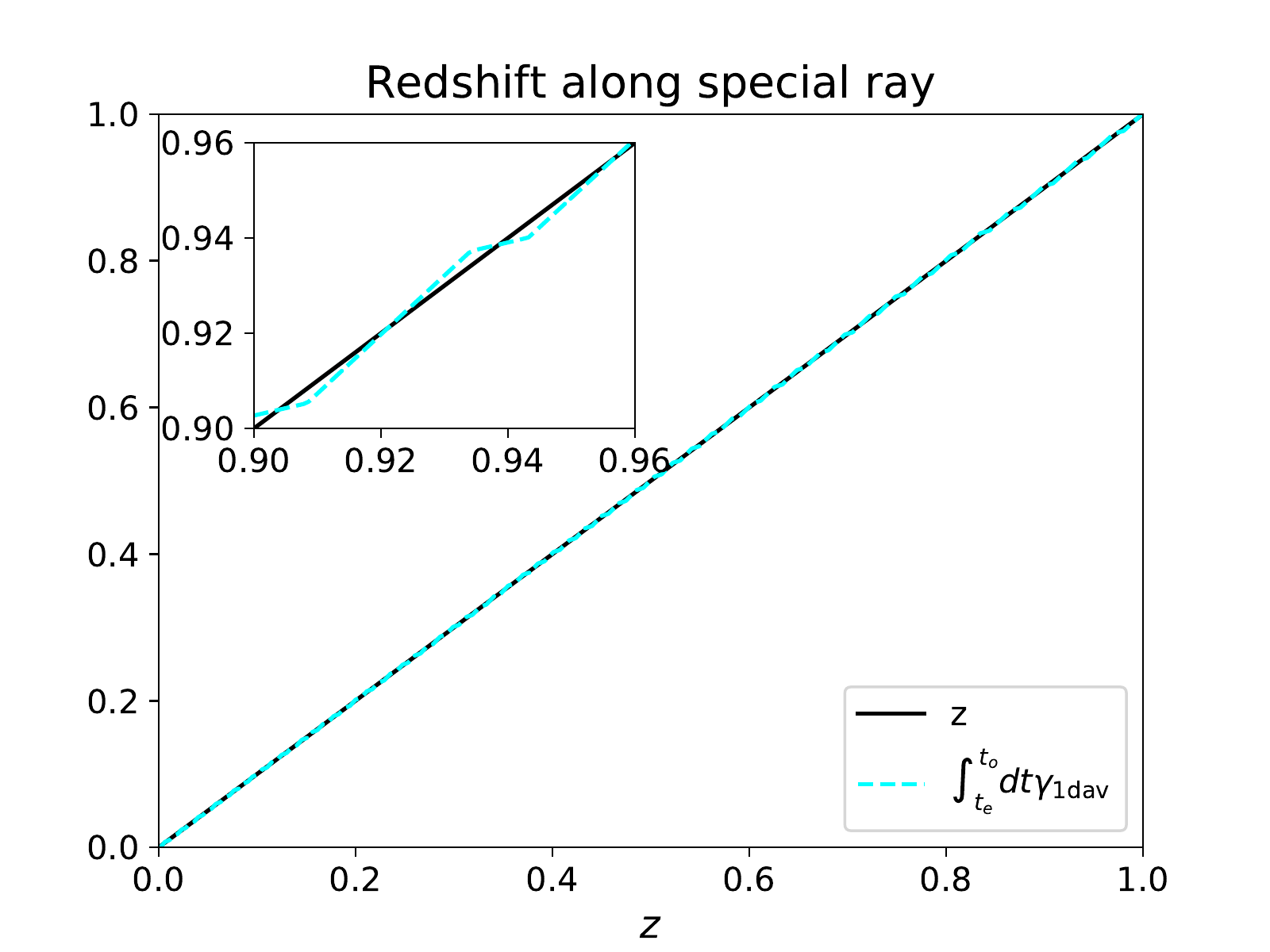}
	\caption{The redshift along a single light ray. The redshift is compared with the ``average'' expectation based on Eq. \ref{eq:z_gamma}. The two lines are indistinguishable in the main plot so a close-up is included to show the fluctuation of the two expression about one another. Notice that the x-axis is the exact redshift which explains why it is the ``average'' expectation which fluctuates.}
	\label{fig:z_special_hellaby}
\end{figure}
The goal with this section is to see whether the mean redshift-distance relation in the Bianch-I tessellated model can be described well by either the Dyer-Roeder approximation or the relation based on spatial averages {\em when there are opaque regions in the model}. Therefore, region 4 (see Table \ref{table:hellaby_model}) is made opaque during light propagation as illustrated in Figure \ref{fig:hellaby_model}. In practice, this is done similarly to what was done in the light propagation study of the N-body simulation. Thus, if a light ray reaches a junction to region 4, it is turned around and propagated the other way.
\newline\newline
One may note that the model studied in this section is not statistically isotropic (but after averaging over many randomly oriented light rays, the model can for a light propagation study be considered effectively statistically isotropic). Therefore, one should not expect that the shear and fluctuations in the expansion rate cancel locally along a light ray. This is indeed not the case, as illustrated in Figure \ref{fig:redshift_hellaby} which shows the redshift components along a single (fiducial) light ray. Instead, a cancellation might be expected if the fluctuations in the expansion rate were computed compared to the spatially averaged expansion rate along the individual light ray. Indeed, consider a light ray propagating in the z-direction, alternately in region 1 and 5. The redshift along this light ray can be written as (using $e^i = k^i/k^t$ and $(k^z)^2 = \left( t/t_0\right)^{-2\gamma}(k^t)^2 $)
\begin{align}\label{eq:z_gamma}
\begin{split}
1+z &= \exp\left(\int_{t_e}^{t_o}dt \left(\frac{1}{3t}\theta + \sigma_{\alpha\beta} e^{\alpha}e^{\beta}\right)  \right) \\
	& = \exp\left(\int_{t_e}^{t_o}dt \frac{\alpha+\beta+\gamma}{3t} + \frac{2\gamma-\alpha-\beta}{3t}\frac{(k^z)^2}{(k^t)^2}\left( \frac{t}{t_0}\right)^{2\gamma}   \right)\\
	& = \exp\left(\int_{t_e}^{t_o}dt \frac{\gamma}{t}  \right).
\end{split}
\end{align}
Thus, the redshift is given by the integral along the light ray of the local ``expansion rate'' in the propagation direction. This is illustrated in Figure \ref{fig:z_special_hellaby} which shows the redshift along a single light ray, where the light ray has been initialized in region 1 in the z-direction and the fluctuations in the expansion rate are measured compared to the 1-dimensional spatial average of $\gamma$, defined as
\begin{align}
	\gamma_{\rm 1dav}:=\frac{g\cdot dz1\left(\frac{t}{t_0} \right) ^{g} + G\cdot dz2\left(\frac{t}{t_0} \right) ^{G}}{dz1\left(\frac{t}{t_0} \right) ^{g}+dz2\left(\frac{t}{t_0} \right) ^{G}},
\end{align}
in analogy with the ordinary (3-dimensional) spatial average defined in the introduction.
\newline\newline
\begin{figure}
	\centering
	\includegraphics[scale = 0.5]{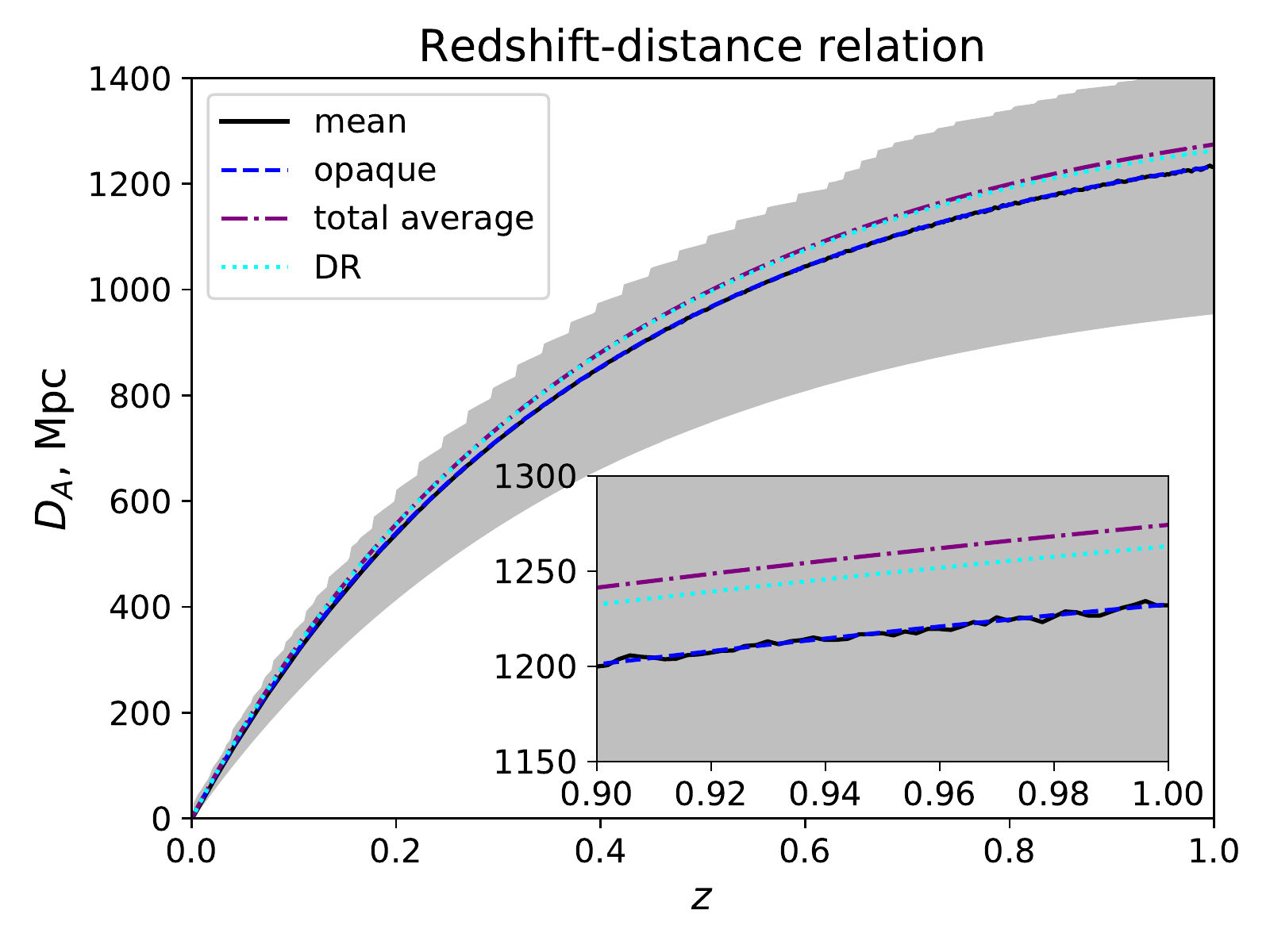}
	\caption{Mean and spread of redshift-distance relation along light rays. For comparison, the redshift-distance relation of the Dyer-Roeder approximation as well as those based on the spatial average of the entire spacetimes (``total average'') and their transparent parts (``opaque'') are also shown. A close-up is included since it is difficult to distinguish all the lines in the main plot. Even in the close-up, it is difficult to distinguish between the mean redshift-distance relation and the line labeled ``opaque''.}
	\label{fig:DAz_hellaby}
\end{figure}
The above discussion highlights that even for the type of model considered in this section, there is a type of local cancellation between fluctuations in the expansion rate and the projected shear along a light ray. However, the cancellation is not identically a cancellation regarding the expansion rate fluctuations about the spatial mean of the entire (transparent part of the) spacetime. Therefore, it is worth studying if the mean redshift-distance relation in this situation will still be well approximated by the Dyer-Roeder approximation if opaque regions are introduced. This is done in the following subsection.

\subsection{Numerical results}
\begin{figure*}
	\centering
	\subfigure{\includegraphics[scale = 0.5]{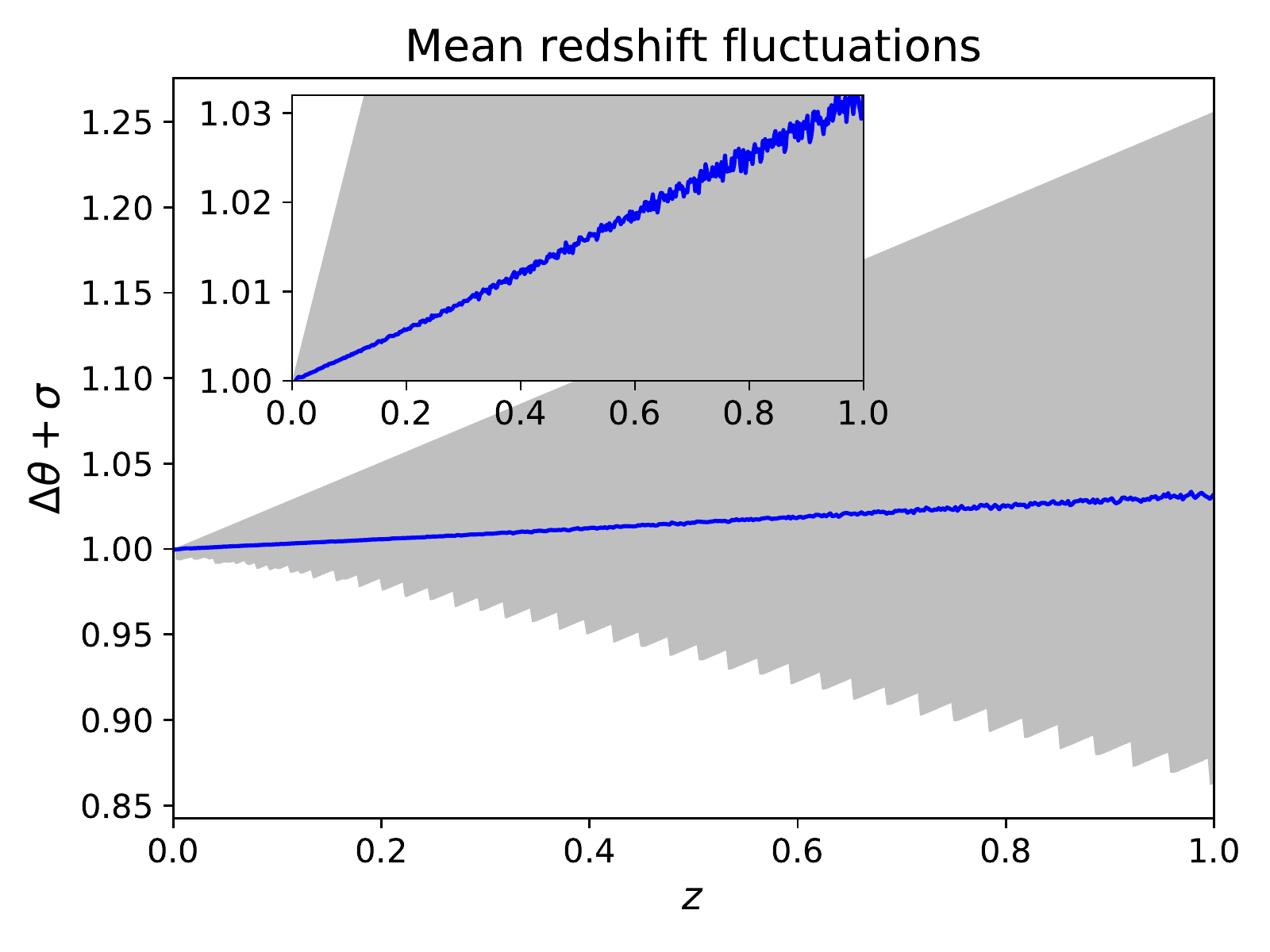}
	}
	\subfigure{\includegraphics[scale = 0.5]{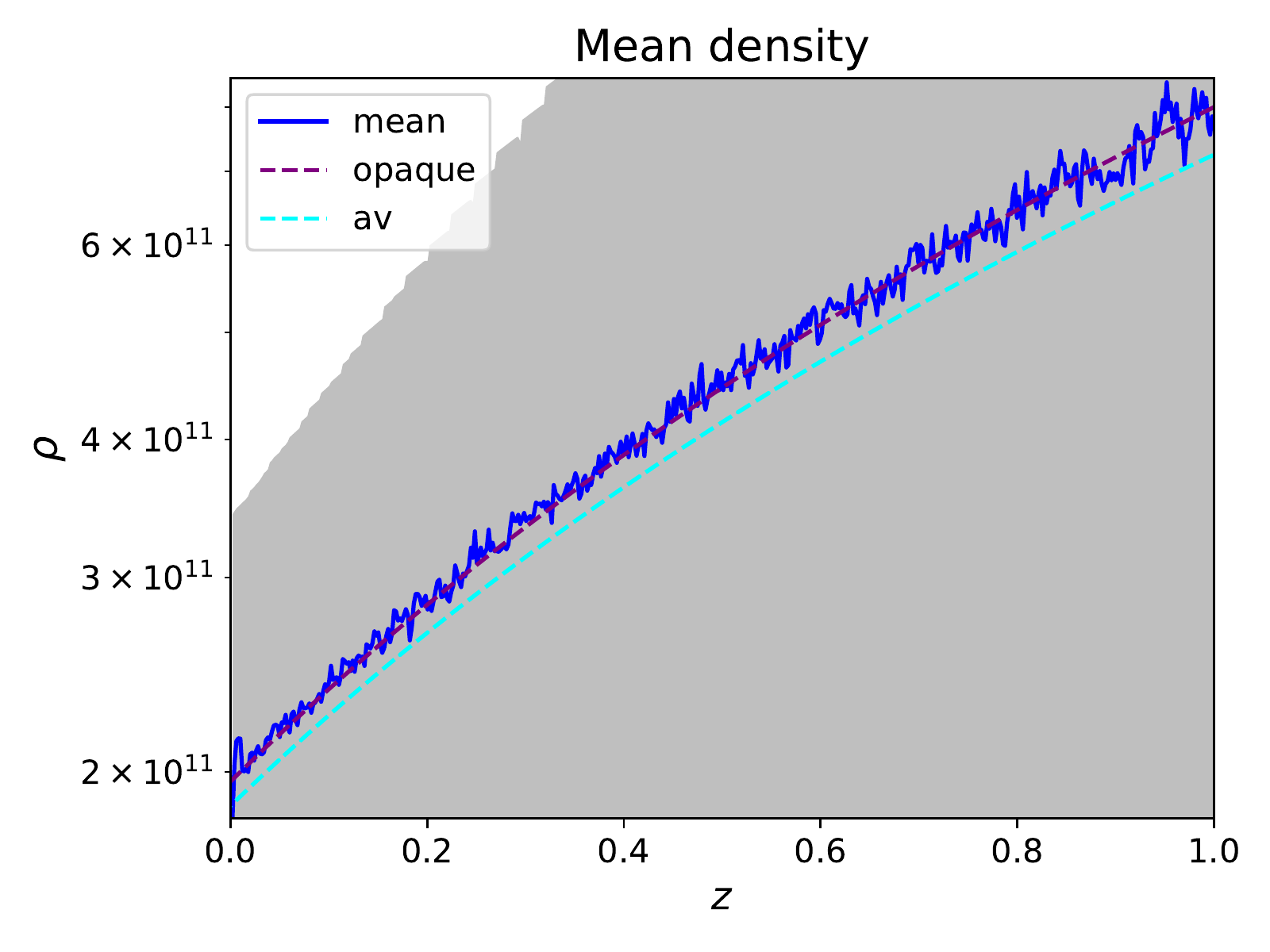}
}
	\caption{Mean and spread of density and fluctuations in the redshift along light rays. The mean density along the light rays is denoted by ``mean'' and is compared to the spatial means based on the entire spacetime (``av'') as well as the spatial average only including the transparent regions (``opaque''). Components of the redshift fluctuations are shown according to $
		\frac{1+z}{1+z_{\rm opaque}} = \exp\left(\int_{t_e}^{t_o}dt \left( \frac{1}{3}\Delta \theta + \sigma^{\alpha}_{\beta}e_{\alpha}e^{\beta} \right)  \right)$, where $z_{\rm opaque}$ is the redshift computed from the spatial average of the expansion rate, only including transparent regions. The fluctuation contribution, $\exp\left(\int_{t_e}^{t_o}dt \left( \frac{1}{3}\Delta \theta + \sigma^{\alpha}_{\beta}e_{\alpha}e^{\beta} \right)  \right)$, is denoted by $\Delta\theta+\sigma$. A close-up of the mean redshift fluctuation is included to highlight the deviation of the mean from 1.}
	\label{fig:zrho_hellaby}
\end{figure*}

The results obtained from propagating 1600 light rays in the model presented in the previous subsection are shown in Figure \ref{fig:DAz_hellaby}. The figure compares the mean redshift-distance relation along the rays with the Dyer-Roeder approximation and the expressions based on spatial averages. Unlike in the previous sections, the Dyer-Roeder approximation is seen to not give a particularly good approximation to the mean. In fact, the prediction based on the Dyer-Roeder approximation deviates very little from the expression based on the spatial averages of the entire spacetime (including region 4). On the other hand, the relation based on spatial averages of the transparent regions gives an excellent prediction of the mean redshift-distance relation.
\newline\indent
The mean density and mean fluctuations in the redshift are shown in Figure \ref{fig:zrho_hellaby}. The figure shows that the mean density along the light rays is well approximated by the spatial mean when not including the opaque regions. Similarly, the mean fluctuations in the redshift compared to the redshift based on the spatial average (again not including opaque regions) are only a few percent.

\section{Conclusions and discussion}\label{sec:conclusion}
It has been argued and illustrated with examples that in standard cosmological scenarios where the Universe can be considered having a global FLRW background, the Dyer-Roeder approximation is valid to the extent that the redshift is well described by the background and the distance is well described by neglecting Weyl lensing and giving the Ricci lensing in terms of the mean density along the light beams. Using explicit examples based on Swiss-cheese models, N-body simulations and results based on perturbation theory, it was argued that the reason the redshift can be well approximated by the background redshift even when there are opaque regions is that the projected shear contribution to the redshift cancels almost exactly with the contribution from the fluctuations in the local expansion rate compared to the background expansion rate, with the cancellation occurring locally along the light rays. With an explicit example based on a somewhat exotic cosmological model which does not contain an explicit FLRW background it was then shown that the Dyer-Roeder approximation is not generally valid. Specifically, in this exotic model the redshift is not well described by that corresponding to the globally averaged spacetime but is instead described well by the redshift corresponding to the average of only the transparent regions.
\newline\indent
While the presented results illustrate the validity of the assumptions upon which the Dyer-Roeder approximation is based in standard cosmological scenarios, it still remains to realistically quantify the mean density along narrow (supernova) light beams in the real universe.

\section{Acknowledgments}
The author thanks Steen Hannestad for assistance with cluster computer facilities.
\newline\indent
The presented numerical results were obtained using computer resources from the Center for Scientific Computing Aarhus as well as UCloud services provided by SDU eScience Center.
\newline\newline
The author is supported by the Carlsberg Foundation.

\end{document}